\documentclass[10pt,conference]{IEEEtran}
 \pdfoutput=1 

\usepackage[utf8]{inputenc}
\usepackage{hyperref}
\usepackage{balance}
\hypersetup{colorlinks,%
            citecolor=black,%
			breaklinks=true,%
            filecolor=black,%
            linkcolor=black,%
            urlcolor=black,%
            pdftex}
\usepackage{url}
\usepackage{amsfonts,amsmath, amssymb} 
\usepackage{booktabs}
\usepackage{multirow}
\usepackage{algorithm}
\usepackage{algpseudocode}
\usepackage{booktabs}
\usepackage{paralist}
\usepackage{comment}
\usepackage{caption}
\usepackage{subcaption}
\usepackage{datetime}

\usepackage{enumitem}
\usepackage{blindtext}
\usepackage{graphicx}
\usepackage{tcolorbox}
\DeclareMathAlphabet\mathbfcal{OMS}{cmsy}{b}{n}
\definecolor{mygray}{gray}{0.5}

\algnewcommand\algorithmicswitch{\textbf{switch}}
\algnewcommand\algorithmiccase{\textbf{case}}
\algnewcommand\algorithmicassert{\texttt{assert}}
\algnewcommand\Assert[1]{\State \algorithmicassert(#1)}%
\algdef{SE}[SWITCH]{Switch}{EndSwitch}[1]{\algorithmicswitch\ #1\ \algorithmicdo}{\algorithmicend\ \algorithmicswitch}%
\algdef{SE}[CASE]{Case}{EndCase}[1]{\algorithmiccase\ #1}{\algorithmicend\ \algorithmiccase}%
\algtext*{EndSwitch}%
\algtext*{EndCase}%

\usepackage{xspace}
\usepackage{xcolor}

\definecolor{darkgreen}{rgb}{0,0.5,0}
\definecolor{brown}{rgb}{0.7,0.3,0}
\definecolor{darkblue}{rgb}{0,0,0.5}
\definecolor{darkred}{rgb}{0.5,0,0}
\definecolor{mygray}{gray}{0.5}

\newcommand{\laurent}[1]{\textit{\textbf{\textcolor{red}{Laurent: #1}}}}
\newcommand{\maria}[1]{\textit{\textbf{\textcolor{darkgreen}{Maria: #1}}}}

\newcommand{\myitem}[1]{\vspace*{0.07in}\noindent\textbf{#1}}

\newcommand{\remove}[1]{}
\newif\ifHideComments

\ifHideComments
	\renewcommand{\aviv}[1]{}
	\renewcommand{\laurent}[1]{}
	\renewcommand{\maria}[1]{}

	\excludecomment{notes}
	\excludecomment{thesiscontents}
\fi

\DeclareMathAlphabet{\mathcal}{OMS}{cmsy}{m}{n}
\newcommand{\system}[0]{{\small\textsf{SABRE}}\xspace}

\begin{document}
\title{\LARGE \bf SABRE: Protecting Bitcoin against Routing Attacks}

\author{\IEEEauthorblockN{Maria Apostolaki}
\IEEEauthorblockA{ETH Z\"urich\\ apmaria@ethz.ch}
\and
\IEEEauthorblockN{Gian Marti}
\IEEEauthorblockA{ ETH Z\"urich\\ gimarti@student.ethz.ch  }
\and
\IEEEauthorblockN{Jan M\"uller} 
\IEEEauthorblockA{ ETH Z\"urich \\jan.m.muller@me.com}
\and
\IEEEauthorblockN{Laurent Vanbever}
\IEEEauthorblockA{ ETH Z\"urich \\lvanbever@ethz.ch} }


\maketitle

\begin{abstract} Routing attacks remain practically effective in the Internet
today as existing countermeasures either fail to provide protection guarantees
or are not easily deployable. Blockchain systems are particularly vulnerable to
such attacks as they rely on working, Internet-wide communication to reach
consensus. In particular, Bitcoin---the most widely-used cryptocurrency---can
be split in half by any AS-level adversary using BGP hijacking.

In this paper, we present \system, a secure and scalable Bitcoin relay network
which relays blocks worldwide through a set of connections that are resilient to routing attacks. \system runs alongside the existing peer-to-peer
network and is easily deployable. As a critical system, \system design is
highly resilient and can efficiently handle high bandwidth loads, including
Denial of Service attacks.

We built \system around two key technical insights. First, we leverage
fundamental properties of inter-domain routing (BGP) policies to host relay
nodes: \emph{(i)} in locations that are inherently protected against routing
attacks; and \emph{(ii)} on paths that are economically-preferred by the
majority of Bitcoin clients. These properties are generic and can be used to
protect other Blockchain-based systems. Second, we leverage the fact that
relaying blocks is communication-heavy, not computation-heavy. This enables us
to offload most of the relay operations to programmable network hardware (using
the P4 programming language). Thanks to this hardware/software co-design,
\system nodes operate seamlessly under high load while mitigating the effects
of malicious clients.

We present a complete implementation of \system together with an extensive
evaluation. Our results demonstrate that \system is effective at securing
Bitcoin against routing attacks, even with deployments as small as 6 nodes.
\end{abstract}


\section{Introduction}
\label{sec:introduction}

Cryptocurrencies, and Bitcoin in particular, are vulnerable to routing attacks
in which network-level attackers (i.e., malicious Autonomous System or AS)
manipulate routing (BGP) advertisements to divert their connections. Once
on-path, the AS-level attacker can disrupt the consensus algorithm by
partitioning the peer-to-peer network. Recent studies~\cite{hijackbtc2017} have
shown that these attacks are practical and can significantly disrupt the
cryptocurrency. Specifically, \emph{any} AS-level attacker can isolate
\raise.17ex\hbox{$\scriptstyle\sim$}50\% of the Bitcoin mining power by
hijacking less than 100 prefixes~\cite{hijackbtc2017}. Such an attack can lead to significant revenue loss for miners and enable exploits such as
double spending.

\myitem{Problem} Protecting against such partitioning attacks is challenging.
On the one hand, local (and simple) counter-measures~\cite{hijackbtc2017} fail
to provide strong protection guarantees. These include countermeasures such as
having Bitcoin clients monitor their connections (e.g., for increased or
abnormal delays) or having them select their peers based on routing
information. On the other hand, global counter-measures are extremely hard to
deploy. For example, systematically hosting Bitcoin clients in /24 prefixes (to
prevent more specific prefix attacks) requires the cooperation of all Internet
Service Providers hosting Bitcoin clients (which is highly unlikely) and would
increase the size of the routing tables. Even heavy protocol modification such
as encrypting all Bitcoin traffic would not be enough to guarantee the
system's safety against routing attacks as the attacker can still distinguish
the traffic from the headers and create a partition.

\myitem{Our work} In this paper we address the fundamental shortcomings of
existing counter-measures. Specifically, we aim at developing techniques that
can secure a system like Bitcoin against routing attacks in a way which:
\emph{(i)} provides strong security guarantees; \emph{(ii)} is partially
deployable, i.e., it should minimize the involvement of third parties;
\emph{(iii)} provides security benefits early-on in the deployment.

\myitem{\system: A Secure Relay Network for Bitcoin} We present \system, a secure relay network which runs alongside the existing Bitcoin network and 
which can protect the vast majority of the Bitcoin clients against routing 
attacks. \system is partially deployable and starts to be useful with as little
as two relay nodes. \system provides strong security guarantees to any connected client---without increasing its load---by enabling it: (i) to learn the latest mined blocks; and (ii) to propagate blocks network-wide, which is essential for miners. We built \system based on two key insights:

\myitem{Insight \#1: Hosting relays in inherently safe locations} The first
insight is to host \system relay nodes in locations: \emph{(i)} that prevent attackers from diverting relay-to-relay connections, guaranteeing \system network integrity; and \emph{(ii)} that are in paths which are---from a routing viewpoint---attractive to many Bitcoin clients, thus reducing the likelihood that attackers will be able to intercept connections to the relay network. To this end, we leverage a fundamental characteristic of BGP policies, namely, that connections established between two ASes which directly peer with each other and which have no customers cannot be diverted. Only such ASes are considered for relay locations. Some of these candidates are also well-connected, making their advertisements attractive. 
Through a thorough measurement study (using real routing data), we show that
such safe locations are plentiful in the current Internet with 2000 ASes being
eligible. These ASes include large cloud providers, content delivery networks, and Internet eXchange Points which already provide
hosting services today and therefore have an incentive of hosting \system
nodes, e.g., for a fee. We also show that 6 \system
nodes are enough to protect 80\% of the clients from 96\% of the AS-level
adversaries (assuming worst case scenario for \system). 
\noindent\textbf{Insight \#2: Resiliency through soft/hardware co-design} As a
publicly-facing and transparent network designed to protect Bitcoin, \system is an obvious target for attackers who could, among others, craft (D)DoS attacks against its publicly-known nodes to disrupt it.
The second insight behind \system is to leverage the fact that: \emph{(i)} the content (blocks) that the relays need to propagate at any given moment in time are predictable and small in size; and that \emph{(ii)} most of the relay operations are communication-heavy (propagating information around) as opposed to being computation-heavy. These two facts enable us to use caching and offload most operations to hardware, in particular, to programmable network devices. This software/hardware co-design enables \system relay nodes to sustain Tbps of load even when originating by malicious actors (DDoS attackers).

We show that our relay node design is practical by implementing in the
P4 programming language~\cite{bosshart2014p4, p4_16_spec}, the default language
for programming network data planes, alongside the (UDP-based) client-side
protocol. Our experiments indicate that P4 is general enough to support \system
and the memory requirements are within the capabilities of today's switches.

\myitem{Contributions} Our main contributions are:

\begin{itemize}
\setlength\itemsep{4pt}

\item The design of \system, a novel relay network that prevents AS-level adversaries from partitioning it (Section~\ref{sec:overview}). 
\item An algorithm for positioning \system nodes in cherry-picked ASes in order to
optimize the security guarantees they provide to the system they aim at
protecting, in our case Bitcoin (Section~\ref{sec:routing}).

\item A novel software-hardware co-design for \system relay nodes that enables them to operate under high load, with minimum software intervention (Section~\ref{sec:node_design}).

\item A thorough measurement study showing the effectiveness of \system in protecting Bitcoin clients. In contrast, we show that existing relays networks provide no protection (Section~\ref{sec:net_eval}).

\item A complete implementation of \system, including the P4 code to run on programmable network switches~\cite{tofino} along with an extension to the Bitcoin client code enabling it to connect to a \system node (Section~\ref{sec:node_eval}).

\item An analysis of the incentives for candidate ASes to host \system nodes
(Section~\ref{sec:deployment}). Among others, we show that eligible candidate
ASes already include well-known cloud providers, which already provide
hosting services.
\end{itemize}

\myitem{Generality} Although \system focuses on Bitcoin, which is by far the most
successful cryptocurrency to date, its routing and system design principles can
be applied to protect any other Blockchain systems whose connections are
publicly routed over the Internet (including permissioned and encrypted ones)
from routing attacks. We discuss the broader applicability of \system in
Section~\ref{sec:discussion}.
\section{Background}
\label{sec:background}

In this section, we first present an overview of BGP and how it can be misused
to perform routing attacks (Section~\ref{ssec:background_bgp}). We then briefly
introduce Bitcoin and the concept of relays (Section~\ref{ssec:background_bitcoin}).

\subsection{Border Gateway Protocol (BGP)}
\label{ssec:background_bgp}

The Internet consists of over 60k individual networks known as Autonomous
Systems (ASes), which rely on BGP~\cite{rfc4271} to exchange information about
how to reach 700k+ IP prefixes~\cite{cidr_report}. Each AS originates
one or more IP prefixes which are then propagated AS-by-AS.

\myitem{Policies} BGP is a single-path and policy-based protocol. Each AS
selects one single best route to reach any IP prefix---including self-owned ones---that it selectively exports to its neighboring ASes (omitting the AS from which it learned the route). These selection
and exportation processes are governed by the business relationships each AS maintains with its neighbors. The most common business relationships are known as \emph{customer-provider} and \emph{peer-peer}~\cite{gao2001stable}. In a
customer-provider relationship, the customer AS pays the provider AS
to get full Internet connectivity. The provider provides such connectivity
by: \emph{(i)} exporting to the customer all its best routes; and \emph{(ii)} exporting the prefixes advertised by the customer to all its neighbors. In a peer-peer relationship, the two ASes connect only to transfer traffic between their respective customers and internal users. They therefore only announce their own prefixes, and the routes learned from their customers to each other. 
Regarding route selection, an AS prefers customer-learned routes over peer-learned ones and peer-learned routes over provider-learned ones. If multiple equally-attractive routes exist (e.g., if two customers announce a route to the same prefix), an AS favors the route with the minimum AS path length towards the prefix before relying on some arbitrary tie-break~\cite{rfc4271}.

\myitem{Hijack} BGP routers do not validate route advertisements. Any malicious
AS can create fake advertisements, known as \emph{BGP hijacks}, for any prefix,
and advertise them to its neighbors. Hijacks constitute an effective way
for an AS to redirect traffic directed to given destinations.

We distinguish two types of hijacks according to whether the fake
announcement contains: \emph{(i)} a more-specific (longer) prefix than a
legitimate one; or \emph{(ii)} an existing (equally specific) prefix. In the former
case, the hijacker AS will attract \emph{all} the traffic addressed to the
more-specific prefix, independently from its position in the Internet topology.
This is because routers forward traffic according to the most specific matching prefix. In the latter case, the rogue advertisement will compete
with the legitimate one. The amount of diverted traffic then depends
on the relative positions of the attacker and the victim in the Internet topology.

\begin{figure}[t]
\begin{subfigure}{.48\columnwidth}
\includegraphics[width=\columnwidth]{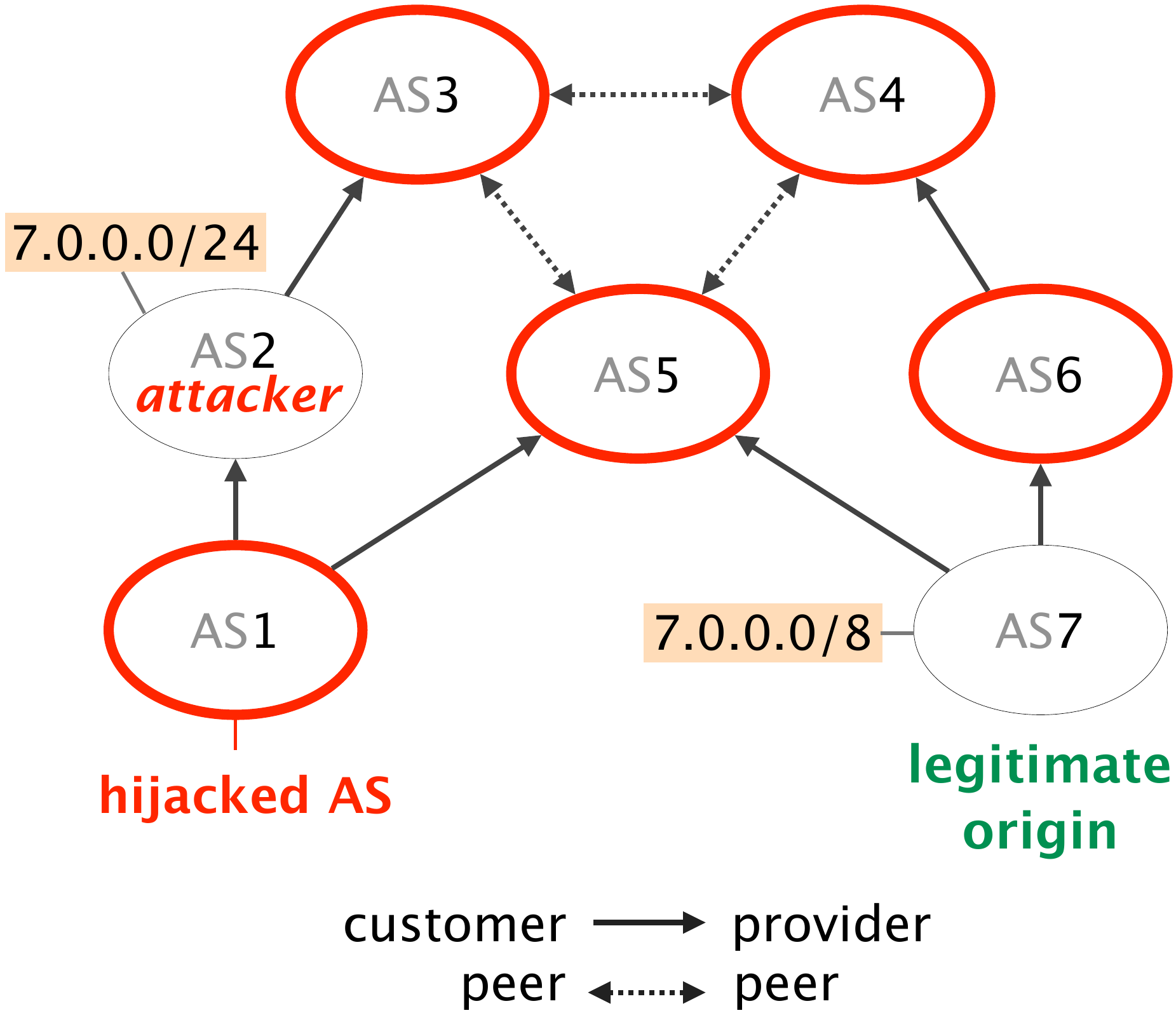}%
\caption{AS 2 hijacks \textsf{7/24}}%
\label{fig:bgp_background_longer}%
\end{subfigure}\hfill
\begin{subfigure}{.48\columnwidth}
\includegraphics[width=\columnwidth]{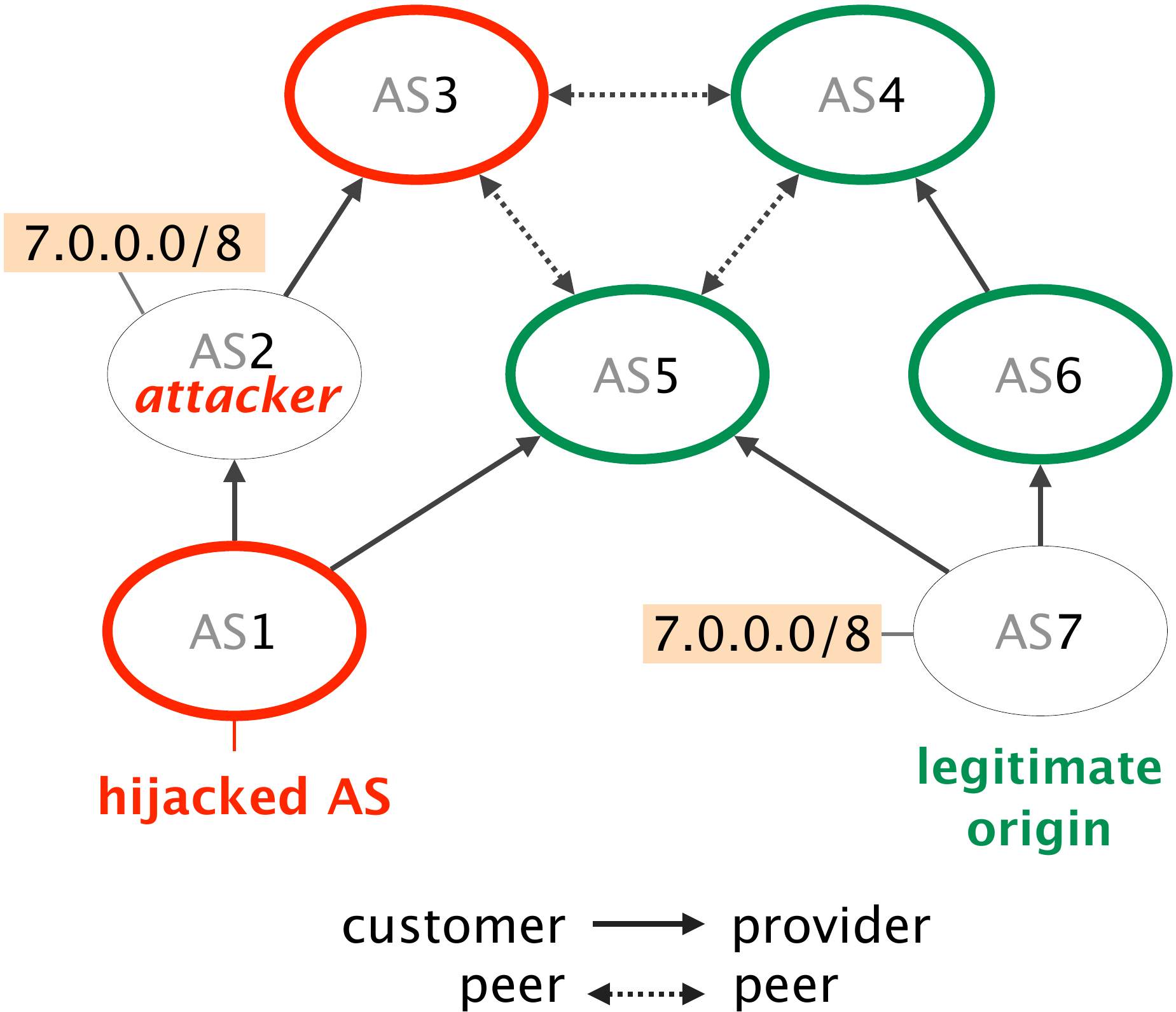}%
\caption{AS 2 hijacks \textsf{7/8}}%
\label{fig:bgp_background_same}%
\end{subfigure}
\caption{The effectiveness of a malicious AS in diverting traffic using BGP hijacks depends on its position and on whether it originates existing prefixes or longer ones. Here, AS 2 attracts traffic from all ASes when originating \textsf{7/24} (a), but only from AS 1 and 3 when originating \textsf{7/8} (b).}
\label{fig:bgp_background}
\end{figure}

Fig.~\ref{fig:bgp_background_longer} illustrates an example of a more-specific
attack in which AS 2 advertises \textsf{7/24}, a more-specific prefix of \textsf{7/8} which is advertised by AS 7. In doing so, AS 2 effectively redirects the corresponding traffic from \emph{all} ASes except AS 7.\footnote{Traffic from AS7 itself is not redirected as AS7 relies on preferred internal
routing protocols such as OSPF to reach its own prefixes internally.} In
contrast, Fig.~\ref{fig:bgp_background_same} illustrates the effect of AS 2
advertising \textsf{7/8} alongside AS 7. AS 2 only manages to attract the
traffic from AS 1 and AS 3. Indeed, AS 1 learns two routes to
\textsf{7/8} from its two providers (AS 2 and 5) and prefers the
illegitimate one from AS 2 because it is shorter. Similarly, AS 3 prefers to reach
\textsf{7/8} using the customer route via AS 2 to the legitimate peer
route learned via AS 5.

More-specific hijacks are more powerful but come with drawbacks. First, such
attacks are more visible since the hijacked prefixes propagate Internet-wide.
In contrast, existing prefixes propagate in smaller regions~\cite{goldberg2010secure}. For
instance, in Fig.~\ref{fig:bgp_background_same}, while AS 4 and AS 5 learn
about the hijacked prefix, they do not propagate it further as they prefer the
legitimate announcement over it. Second, network operators often filter BGP advertisements whose prefixes are longer than /24~\cite{Karlin:2006:PGB:1317535.1318378}, thus preventing
more-specific attacks against existing /24. Of course, an attacker can still
advertise the /24, i.e., an existing prefix.

By default, hijacking a prefix creates a black hole at the attacker's location.
However, the attacker can turn a hijack into an \emph{interception} attack and make 
herself a man-in-the-middle (MITM) by preserving at least one
path to the legitimate origin~\cite{defcon_mitm,
goldberg_how_secure_are_interdomain_routing_protocols}. For instance, in
Fig.~\ref{fig:bgp_background_longer}, AS 2 could selectively announce
\textsf{7/8} to AS 1 to keep a working path to the legitimate origin via
AS 3. Observe that AS 2 cannot achieve the opposite interception attack, i.e.,
divert the traffic from AS 3 and redirect it to AS 1 instead, as it does
not learn a path to the legitimate origin via AS 1.

\subsection{Bitcoin}
\label{ssec:background_bitcoin}

Bitcoin is a decentralized transaction system which relies on a randomized peer-to-peer
network to implement a replicated ledger, the \emph{Blockchain}, which keeps track of the ownership of funds and the balance of each Bitcoin
address. The Bitcoin network disseminates two types of information:
\emph{transactions} and \emph{blocks}. Transactions are used to transfer value
from one address to another, while blocks are used to synchronize the state of
the system. Bitcoin peers are identified by their IP address, connect to each other using TCP, and exchange data in plain text. Bitcoin comprises around 10k publicly reachable peers~\cite{bitnodes_website}, and 10$\times$ more NATed peers~\cite{Biryukov:2014:DCB}.

Blocks are created by \emph{miners} and contain the latest transactions as well
as a Proof-of-Work (PoW). A PoW is a computationally-heavy puzzle, unique for
every new block, whose difficulty is regularly adapted such that it takes 
10 minutes on average to generate a new block~\cite{nakamoto2008bitcoin}. A newly mined block is  propagated network-wide and included in the blockchain according to consensus, thereby yielding a financial reward to its miner.
Bitcoin participants unaware of the latest blocks will waste their mining power and can be fooled into accepting invalid transactions. 

\myitem{Relay networks} are overlay networks maintained by a single administrative entity which run alongside Bitcoin's peer-to-peer network. Relay networks aim at assisting the Bitcoin network, not replacing it. As an illustration, the three most well-known relays nowadays are: Falcon~\cite{falcon}, FIBRE~\cite{fibre}, and the Fast Relay Network
(FRN)~\cite{corallo} and aim at speeding up block propagation. These relay networks rely on a system of high-speed relay nodes and/or on advanced routing techniques. By connecting to these relays, a client can alleviate the effects of bad network performance that may otherwise affect the time needed to acquire a new block.

\section{\system: A Secure Relay Network for Bitcoin}
\label{sec:overview}

\system is a transparent relay network protecting Bitcoin clients from routing
attacks by providing them with an extra secure channel for learning and
propagating the latest mined block. By transparent, we mean that the IP addresses of the \system relay nodes will be publicly known (e.g. via a website) and that every Bitcoin client is welcome to connect to them. To benefit from \system, a Bitcoin client simply needs to successfully establish a
connection to at least one relay node. \system relays contribute to block propagation by receiving, validating and transmitting new blocks to all connected clients.

To achieve its goals, the \system network \emph{must} remain connected at all
times, even under arbitrary routing or DDoS attacks. \system leverages two key
insights to guarantee connectivity: \emph{(i)} smart
positioning of the relays to secure its internal connections and minimize the
clients attack surface (Section~\ref{ssec:routing}); and \emph{(ii)} a
hardware/software co-design to enable relay nodes to sustain almost arbitrary
load (Section~\ref{ssec:system}). We note that these insights can be used to secure other Blockchain systems against routing attacks (Section~\ref{sec:discussion}). We start by describing our attack model (Section~\ref{subsec:model}).

\begin{figure*}[t]
\centering
\begin{subfigure}{.33\textwidth}
 \includegraphics[width=\columnwidth]{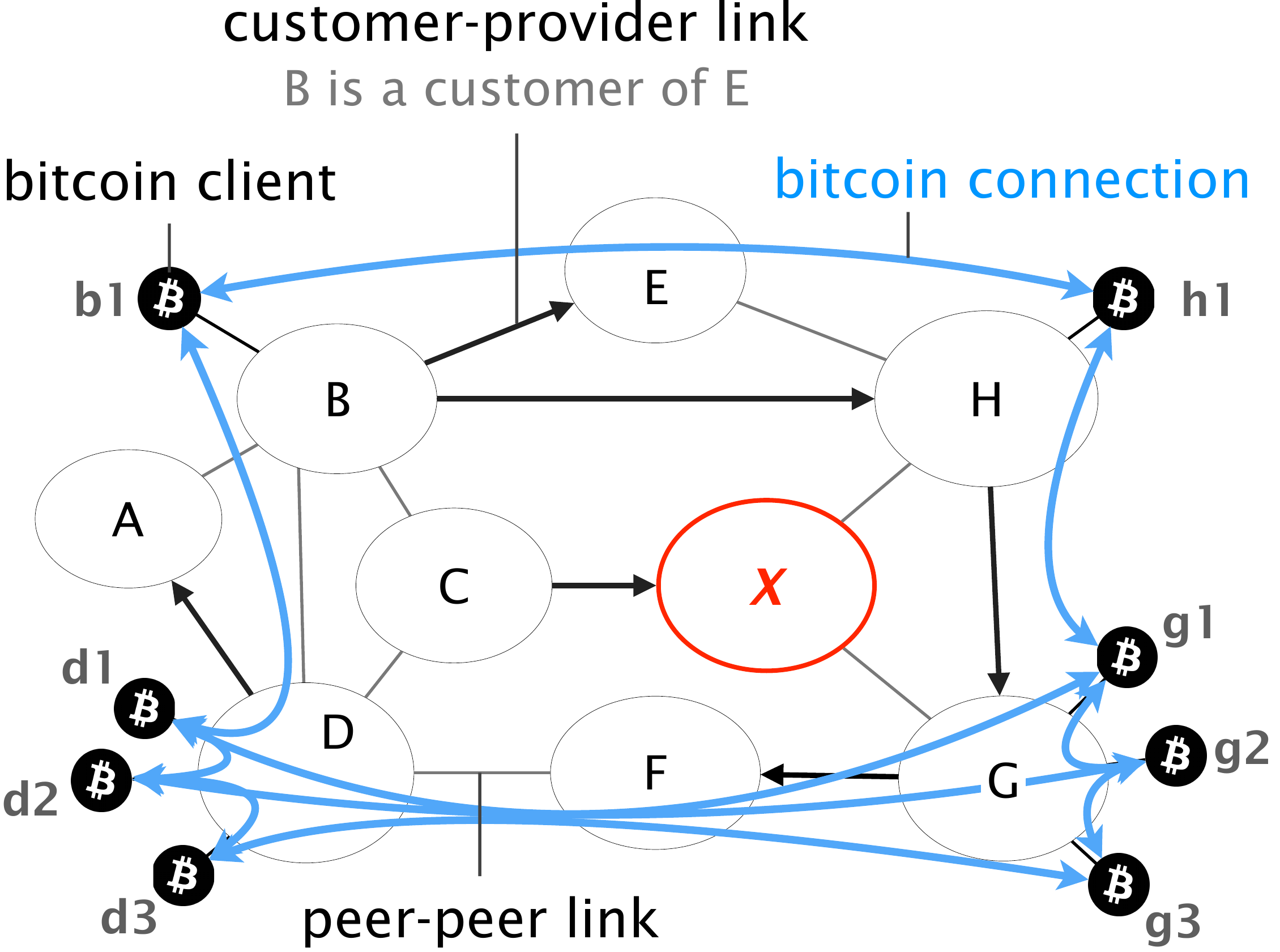}
 \caption{AS-level topology}
 \label{fig:att1}
\end{subfigure}\quad%
\begin{subfigure}{.31\textwidth}
\includegraphics[width=\columnwidth]{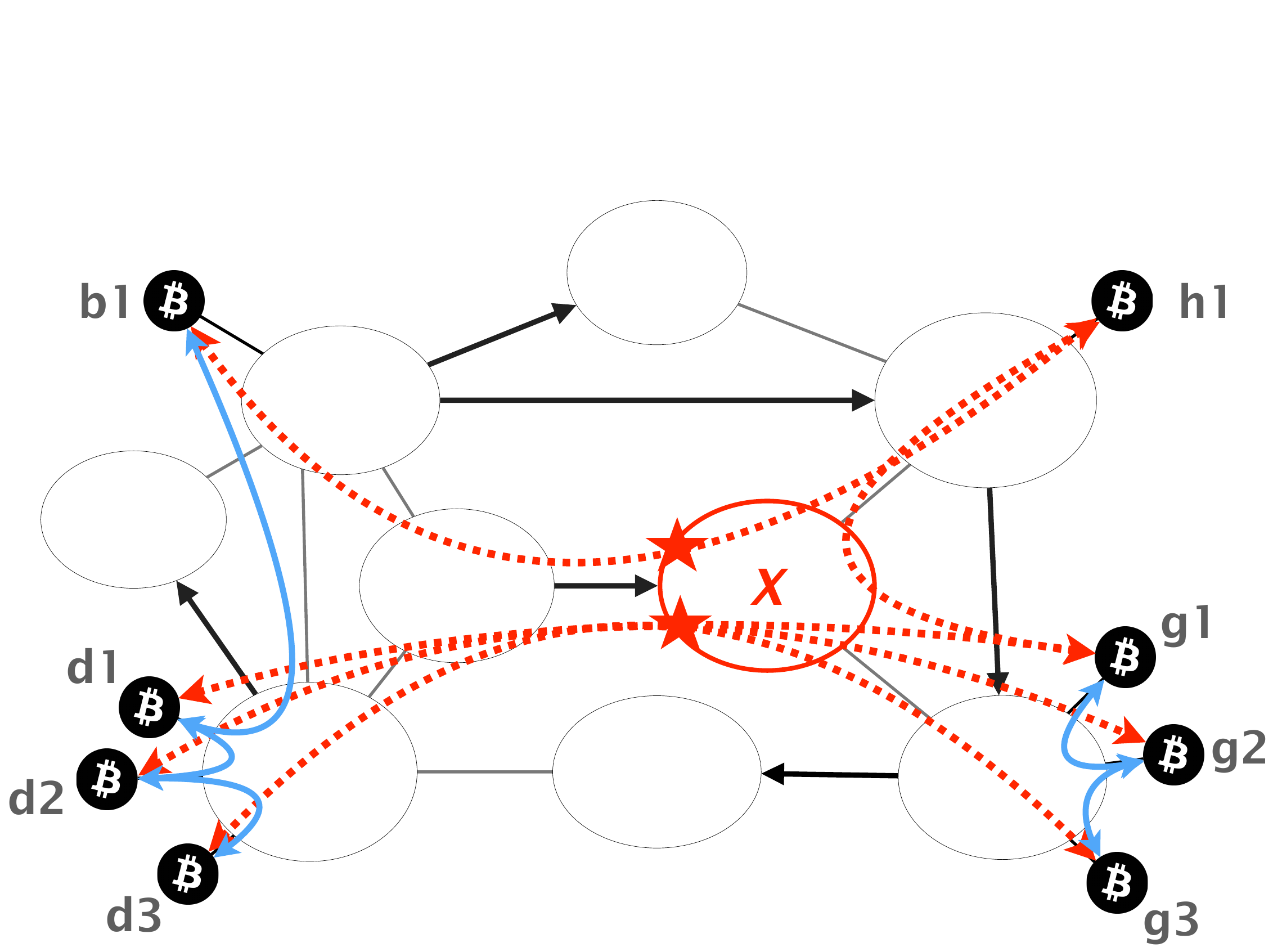}%
\caption{AS X hijacks ASH \& ASG}%
\label{fig:att2}%
\end{subfigure}\quad%
\begin{subfigure}{.31\textwidth}
\includegraphics[width=\columnwidth]{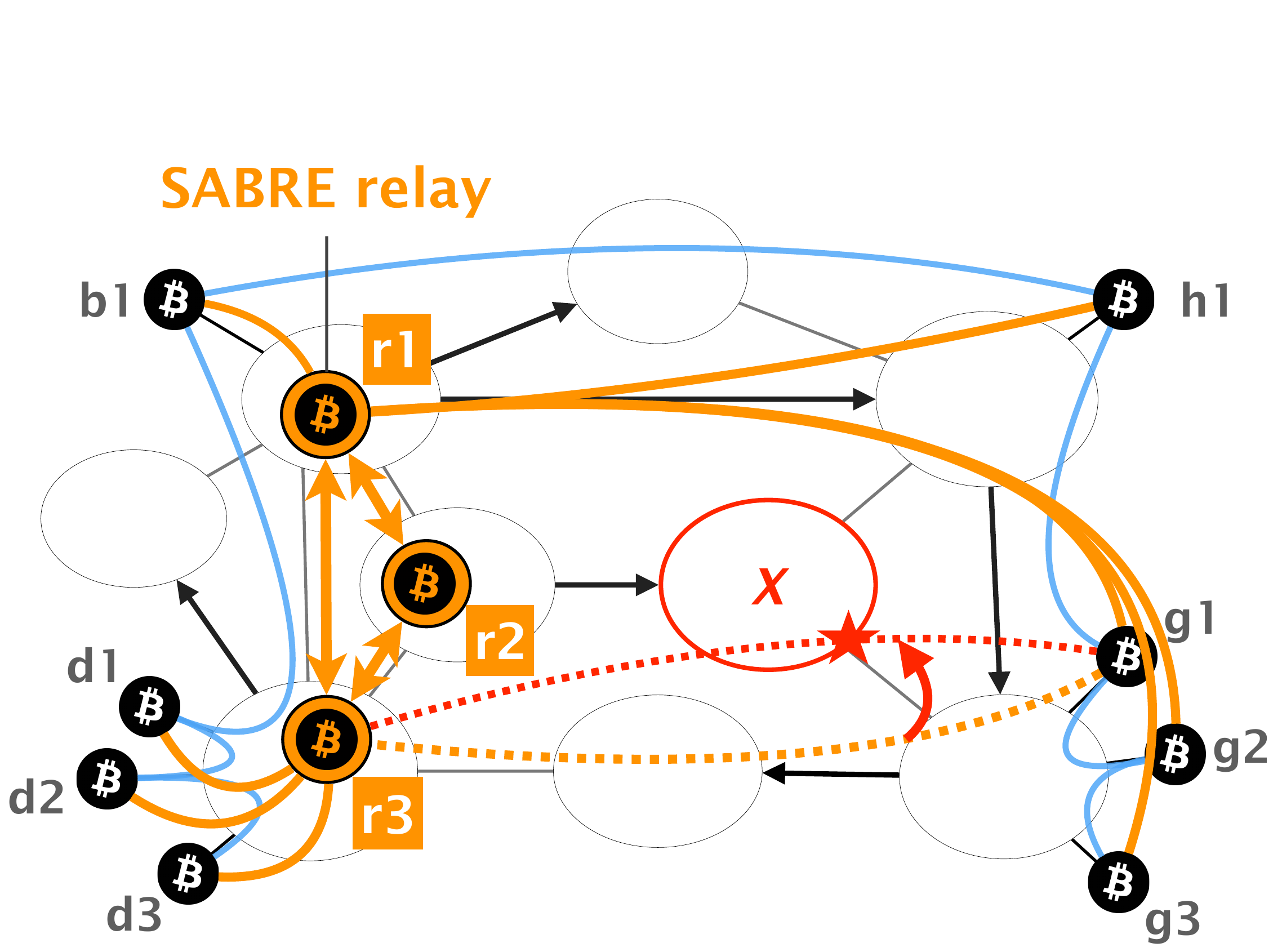}%
\caption{With \system, network stays connected}%
\label{fig:att3}%
\end{subfigure}%
\caption{\system protects the Bitcoin network from AS-level adversaries aiming to partition it. Without \system, AS X can split the network in half by first diverting traffic destined to AS H and AS G using a BGP hijack and then dropping the corresponding connections (Fig.~\ref{fig:att2}). With \system, the network stays connected (Fig.~\ref{fig:att3}).
}
\label{fig:mainfig}
\end{figure*}

\subsection{Attacker Model}
\label{subsec:model}

We consider a single AS-level attacker whose goal is to partition the Bitcoin network into two disjoint components $S$ and $N$. To do so, she first diverts the traffic destined to $S$ or $N$ by performing an interception attack using existing and more-specific prefixes (Section~\ref{sec:background}). The attacker then: \emph{(i)} identifies the Bitcoin connections by inspecting the network and/or transport layer headers (i.e., by matching on IP addresses and/or TCP/UDP ports); and \emph{(ii)} drops the connections bridging the partition. 
Such an attacker is powerful and can effectively split the Bitcoin network into half~\cite{hijackbtc2017}. In fact, partitioning any Blockchain system constitutes an effective DoS attack, and can result in revenue loss and allows double spending (see Section~\ref{sec:discussion}). 

We assume that the attacker knows: \emph{(i)} the IP addresses of \emph{all} \system nodes; along with \emph{(ii)} the code running on the relay nodes. As such, the attacker can also hijack the prefixes hosting relay nodes and drop \emph{all} traffic destined to them. Alternatively, the attacker can  perform a DDoS attack on the relay network, by directing load against its nodes aiming at exhausting their resources.

\noindent\textbf{Example} We illustrate the attack using Figure~\ref{fig:att1} and
\ref{fig:att2} which depict a simple AS-level topology composed of 9 ASes. $ASB$, $ASD$ $ASH$ and $ASG$ host Bitcoin clients who
establish Bitcoin connections to each other (in blue). $ASX$ is
malicious and aims at disconnecting the nodes on the left side
($S=\{b1,d1,d2,d3\}$) from the others ($N=\{h1,g1,g2,g3\}$). To that end,
$ASX$ intercepts the Bitcoin connections directed to
$N$ by hijacking $ASH$ and $ASG$ prefixes. As a result, $ASX$
diverts all the connections from $S$ to $N$, and some more (e.g., the
connection from $h1$ to $g1$). We depict the diverted connections in red in Figure~\ref{fig:att2}. Once on-path, the attacker drops the Bitcoin traffic
\emph{crossing} the partition and forwards the rest normally. For instance, the
attacker does \emph{not} drop the connection between $h1$ and $g1$ and simply
relays it from $AS H$ to $ASG$ untouched. Once the attack is launched, nodes in
$S$ can no longer communicate with nodes in $N$: the Bitcoin network is partitioned.

\subsection{\system secure network design}
\label{ssec:routing}
We now explain the routing properties of \system node placement that protect relay-to-relay, client-to-relay, and relay-to-client connections from being disconnected. We describe an algorithm for systematically finding such locations in Section~\ref{sec:routing}.

\myitem{Protecting relay-to-relay connections} \system network is composed of
relay nodes that are hosted in /24 prefix in ASes that: \emph{(i)} have no
customers; \emph{(ii)} have direct peering connections; and \emph{(iii)} form a $k$-connected graph. 

These constraints secure relay-to-relay connections from routing attacks for four reasons.
\emph{First}, they prevent any attacker from diverting traffic among relays by
advertising a more-specific prefix, forcing her to compete with legitimate advertisements, namely to advertise existing prefixes. Indeed, routers will discard any advertised prefix that is longer than $/24$. \emph{Second}, these constraints prevent any attacker from advertising an economically strictly \emph{better} route
than the legitimate one. This is because the ASes hosting relay nodes learn the legitimate prefixes via direct peer links and do not have customers, meaning that no malicious AS can advertise a more preferable route. \emph{Third}, these constraints limit the number of malicious ASes which can advertise an equally-preferable route to only those ASes that directly peer with the ASes hosting relay nodes. 
Finally, they ensure that the chances for such attackers to divert relay connections decrease exponentially as $k$ (i.e., the connectivity of the relay graph) increases. Indeed, BGP routers rely on an arbitrary tie-break to select among equally-preferred routes
(e.g. by choosing routes learned from the lowest peer address~\cite{rfc4271}). Assuming that the attacker is equally likely to win this tie-breaking, she would only have a 3.1\% ($0.5^5$) probability of disconnecting a 5-connected relay network. In Section~\ref{sec:net_eval}, we show that well-connected relay networks are numerous.

\remove{
In this case, the attacker's advertisements could only be more preferred if the attacker was a
customer of the relay AS, which contradicts constraint \emph{(i)}. Indeed, if the attacker is not directly connected its advertisement will not be preferred
 \emph{Second}, only ASes that directly peer with the relay ASes stand a chance to 
 \emph{Third}, being $k$-connected, disconnecting
the \system network would require $k$ cuts (constraint \emph{(iii)}). As
such, it will not be affected by hijacks from peers unless they are at least $k$.

}
\remove{
\emph{Second}, if the hijack comes from a peer, it can be precisely detected
with no false positives. Indeed, a fake advertisement will stand out due to the
origin being the attacker AS. Likewise, if the attacker advertises a longer
path (e.g. by prepending the legitimate AS to her advertisement), it would be
less preferred than the original one, which is received over a direct peering
link (constraint \emph{(ii)}).
}

\myitem{Protecting client-to-relay connections} While we can selectively place
relay nodes (we discuss the incentives for ASes to host relay nodes in Section~\ref{sec:deployment}), we cannot re-position all Bitcoin clients (or host them in /24 prefixes). This means that client-to-relay connections cannot be made inherently secure against \emph{all} possible AS-level adversaries and active routing attacks.

We protect client-to-relay connections by further restricting where
we host \system relays to ensure that they not only meet the criteria to secure
relay-to-relay connections but also that their respective advertisements will
tend to be preferred by ASes with Bitcoin clients over competing ones. Doing so we can lower the
amount of traffic a malicious AS can effectively divert, i.e. maximize \system's coverage.

Although, individual relays are unlikely to protect an AS against all possible
attackers, a set of relays can often do so. Indeed, this can happen if for each
possible AS-level adversary there is a relay who can beat her by offering a
better route. As the set of Bitcoin clients tend to be highly centralized~\cite{hijackbtc2017}, we show in Section~\ref{sec:net_eval} that a relatively small relay network is enough to protect many clients.

\myitem{Protecting relay-to-client connections} Finally, an attacker might try to attack traffic sourced by the relay network \emph{to} the Bitcoin clients. For instance, an attacker could hijack the prefixes of Bitcoin clients and drop the relay connections by matching on any relay IP address. While this attack is more cumbersome (there are way more clients than relays), it is nonetheless possible. \system prevents this attack by obfuscating the traffic exchanged between the clients and the relay nodes, forcing the attacker to perform full inspection (beyond L4 headers) on a possibly huge volume of diverted traffic, henceforth rendering the attack highly impractical compared to the gain. Observe that while encrypting the already-obfuscated traffic would render even full inspection useless, encryption alone would not help as the attacker would still be able to match on the destination IP.  

To obfuscate the traffic we suggest two techniques. First, the relays could modify their source IP addresses when sending to the regular clients. This is possible as \system uses connectionless communications between the relays and the clients, enabling clients to accept packets with a different source IP than the one they send traffic to. Second, clients
could connect to the relay via a VPN/proxy service, forcing the attacker to first find the mapping between the proxy IP and the corresponding bitcoin client.

\myitem{Example} Using Fig.~\ref{fig:att3}, we now explain how a \system
deployment of three relays, namely $r1$, $r2$ and $r3$, protects against routing
attacks such as the one shown in Fig.~\ref{fig:att2} by securing intra-relay
connectivity and maximizing coverage.

With respect to Fig.~\ref{fig:att1}, each Bitcoin client is now connected to a
least one relay node in addition to maintaining regular Bitcoin connections.
Here, nodes $g_1, g_2, g_3$ are connected to relay $r_1$ while node $g_1$ is
also connected to node $r_3$. Hosted in ASes that peer  directly, relay-nodes protect their internal connectivity against $ASX$'s hijacks. For instance, consider that $ASX$ advertises the $/24$ prefix covering $r1$ to $ASC$. Since $ASX$ is a provider of $ASC$, $ASC$
discards the advertisement as it prefers to route traffic via a peer instead. 
At the same time, forming a 2-connected graph allows the relay network to sustain any single link cut. This can be caused by a failure, an agreement violation or an unfiltered
malicious advertisement from another direct peer such as $ASF$ to $ASD$. Observe that this would not hold if $r_2$ was not deployed.
Finally, the exact positioning of relays is such that the paths towards them are more preferred over those of the attacker.
As an illustration, $ASX$ can divert the connection from $ASG$ to $ASD$ by
advertising a path with a better preference (as a peer) than the one originally
$ASG$ uses (a provider route, via $ASF$). Even so, $ASX$ cannot divert the
connection from $ASG$ to $ASB$. Indeed, $ASG$ will always prefer its customer
path over any peer path.

\remove{
\system deployment is built so as to maximize the
chances that, for every possible attacker and each client, there is an AS with
a relay node which offers a better path to the relay than the attacker's one.
As an illustration, $ASX$ can divert the connection from $ASG$ to $ASD$ by
advertising a path with a better preference (as a peer) than the one originally
$ASG$ uses (a provider route, via $ASF$). Yet, $AS$ cannot divert the
connection from $ASG$ to $ASB$. Indeed, $ASG$ will always prefer its customer
path over any peer path.

In terms of reliability, $ASA$, $ASB$, and $ASC$ form a $2$-connected graph
which enables to sustain intra-relay connectivity under a single link cut. Such
a cut can occur due to a failure, an agreement violation or an unfiltered
malicious advertisement from another direct peer such as $ASF$ to $ASD$.
Observe that this would not hold if $r_2$ was not deployed. \laurent{The
intuition behind $k$-connectivity isn't clear in the overview. Also, we mix
things around saying in some places that intra-relay connections cannot be
hijacked and here saying that they can. need to fix that.}
}

\begin{figure}[t]
 \centering
 \includegraphics[width=.8\columnwidth]{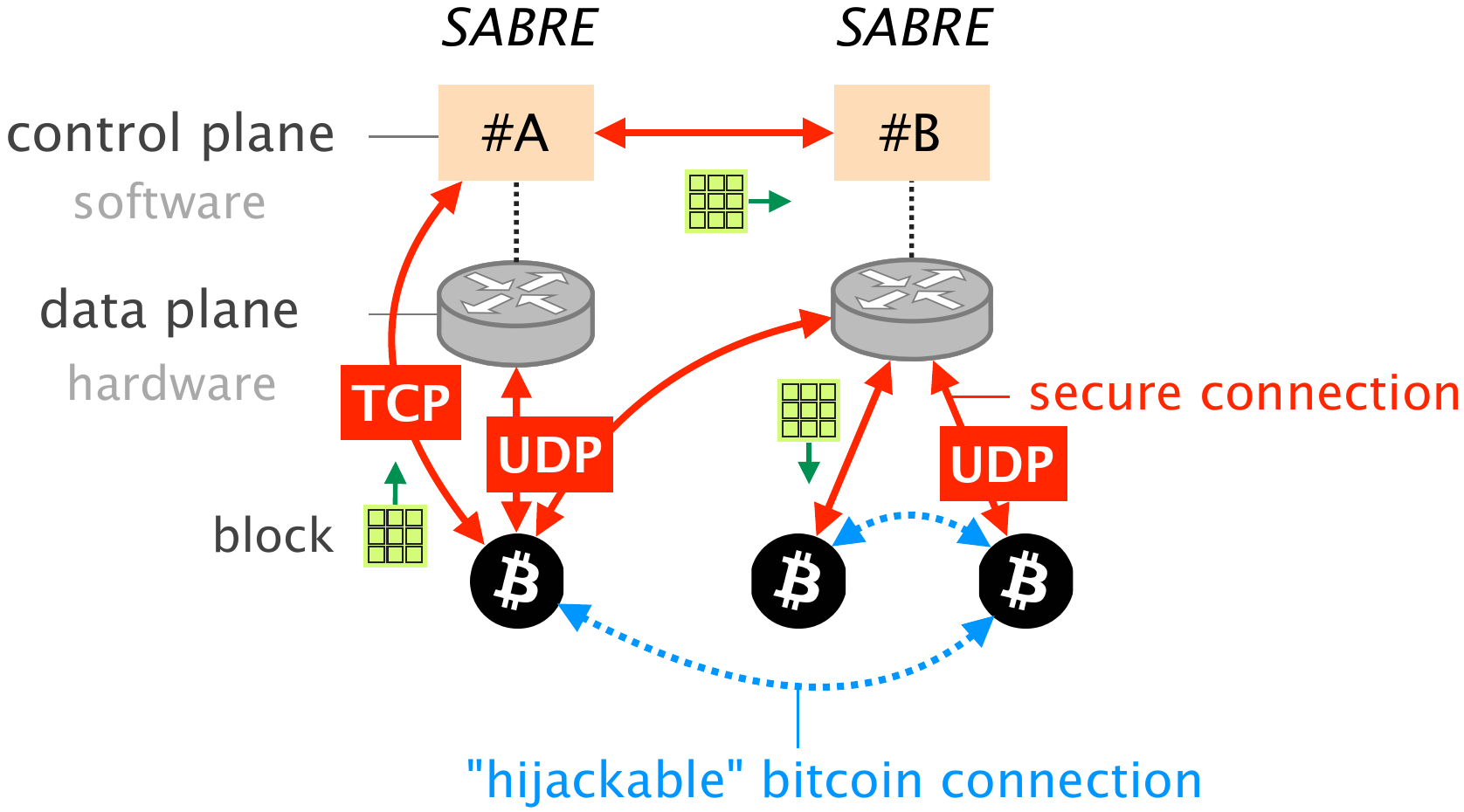}
 \caption{\system offloads most communication to the switch}
 \label{fig:system_design}
\end{figure}

\subsection{\system resilient software/hardware node co-design}
\label{ssec:system}

As a publicly-known and accessible relay network, \system nodes should be able to sustain high load,
either caused by legitimate Bitcoin clients or by malicious ones who try to
exhaust their resources. 
To scale, \system nodes rely on software/hardware co-design in which most of
the operations are offloaded to programmable network switches (e.g., P4-enabled
ones). As an illustration, in Fig.~\ref{fig:system_design} two \system nodes are connected to each other and to some Bitcoin clients. One client talks directly to the controller via the switch, while the others only to the switch.

Particularly, \system's relay design is based on the observations that:
\emph{(i)} the content that needs to be cached in relay node is predictable and small in size, consisting in the one or two blocks of 1MB that were most recently mined; and \emph{(ii)} most of the relay
operations are communication-heavy, consisting in propagating the latest known
block to many clients and distinguishing the new one. The former allows effective caching (extremely high hit rate) while the latter allows for a partially hardware implementation in programmable network devices. This software/hardware co-design enables \system nodes to operate at Tbps and therefore sustain large
DDoS attacks. Indeed, Barefoot Tofino programmable network devices can deal with as much as 6.5 Tbps of traffic in the backplane~\cite{tofino}. 

While using programmable network devices enable high performance, it does not
make it easy due to the lack of a broad instruction set and the strict limitations with respect to memory and number of operations per packet.  
We overcome these limitations with three techniques. First, our
software/hardware design seamlessly blends in hardware and software operations,
enabling to automatically escalate operations that cannot be done in the switch
to a software component. In \system, only the validation of new blocks (which
happen once every 10 minutes) needs to be escalated while all other requests are served by the hardware over a UDP-based protocol. Second, our implementation
relies on optimized data structures which are both memory efficient and require a fixed number of operations per-access. Third, we heavily precompute and cache values that would need to otherwise be computed on the switch (e.g., UDP checksums).
\section{\system Secure Network Design}
\label{sec:routing}

In this section, we formally define the problem statement
of selecting the relay hosting ASes (Section~\ref{subsec:problem}) so as to minimize the
possibility of a successful routing attack against Bitcoin before presenting
an algorithm for solving it in Section~\ref{subsec:algo1} and \ref{subsec:algo2}. 

\remove{
the of any desired size and connectivity.
To that end, we first formally define the problem which can be split into three main sub-problems, namely \emph{(i)} finding candidate AS to host relays; \emph{(ii)} systematically predicting the path that each AS with Bitcoin clients will follow in the presence of a malicious conflicting advertisement of a given relay prefix from all potential AS-level adversaries; and \emph{(ii)} solving an optimization problem to find relay network of $N$ relays and connected $K$ with the maximum effectiveness against AS-level adversaries.
}

\subsection{Problem Statement \& Challenges}
\label{subsec:problem}
The security provided by \system depends on: (i) how secure the intra-relay connectivity is, i.e., how many connections an AS-level adversary needs to hijack to disconnect
the graph; and (ii) how much of the Bitcoin network is covered, i.e., how
likely it is that an AS-level adversary will be able to prevent each particular
client from connecting to \emph{all} relay nodes.

Thus, given a level of intra-relay connectivity to achieve (e.g., $2$-connectivity), our goal is to maximize the Bitcoin coverage. Formally, we define our problem as follows:

\myitem{Problem statement} Let ${G} = (\mathcal{AS},E)$ be the AS-level
topology graph in which vertices ($\mathcal{AS}$) correspond to ASes and edges
($E$) to inter-AS links. Let also $\mathcal{B} \subseteq \mathcal{AS}$ be the
subset of ASes that host Bitcoin clients and $\mathcal{R} \subseteq
\mathcal{AS}$ be the subset of ASes that have no customers. Finally, let $G' =
G[\mathcal{R}, E']$ be the subgraph of $G$ induced by $\mathcal{R}$ and the
subset $E' \subseteq (\mathcal{R} \times \mathcal{R})$ of peer-peer
inter-AS links.
We define $\mathcal{A}=\mathcal{AS} \times \mathcal{B}$ as the set of all
attack scenarios, namely all pairs of ASes $(a,b)$ in which AS $a$
acts as AS-level adversary for AS $b$ with Bitcoin clients. Let $\mathcal{S}: \mathcal{R} \rightarrow \mathcal{A}$ be a function which, given a candidate relay AS, finds the subset $\alpha \subseteq  \mathcal{A}$ of attack scenarios that this candidate AS protects against. Let furthermore $\mathcal{C}: \mathcal{P}(\mathcal{A}) \rightarrow \mathbb{R}$ be a function ($\mathcal{P}(\cdot)$ denotes the power set) which, given a set of attack scenarios $\alpha \subseteq \mathcal{A}$, quantifies their significance for the Bitcoin system by computing 
the sum of all scenarios in $\alpha$ weighted by the number of Bitcoin clients hosted in the victim AS, i.e. $\mathcal{C}(\alpha) = \sum_{(x,v)\in \alpha}^{} w_v  $ where $w_v$ is the number of Bitcoin clients in AS $v$.

We want to find the subgraph $G''=G'[R']$ of $R' \subseteq \mathcal{R}$ such
that $\lvert R'\rvert = N$; $G''$ is $k$-connected; and $\mathcal{C}
\big(\bigcup_{r_i \in R' } \mathcal{S}(r_i) \big)$ is maximized. Put differently, we aim---for a fixed number of relays $N$ and relay inter-connectivity $k$---at maximizing the number of attack scenarios Bitcoin clients are protected
against.

\myitem{Challenges} Solving the above problem optimally is challenging for at
least three reasons. First, the effectiveness of any subset of relays $R'$ 
 depends on the \emph{union} of the sets of the attack scenarios each relay $r \in R'$ protects against. As these are in general not disjoint, this problem reduces to the maximum coverage problem.
Second, finding $k$-connected subgraphs in a random graph is difficult~\cite{kconne}.
Third, one needs to be able to predict the forwarding path from each AS with Bitcoin clients to a relay considering \emph{any} possible attacker.

We develop a heuristic to address the first two challenges (Section~\ref{subsec:algo1}) and an algorithm for finding the possible attack scenarios for every attacker (Section~\ref{subsec:algo2}). 

\subsection{Positioning \system Relays}
\label{subsec:algo1}
Given a number $N$ of relays and their desired connectivity $k$, our
algorithm returns a set of ASes $R'$ in which to host relays such that the
connectivity and size requirements are met and weighted coverage is maximized.
This maps to the maximum coverage problem with an additional connectivity constraint. Thus, we use a greedy approach, shown to be effectively optimal for the maximum coverage problem \cite{feige}.

Our algorithm starts with an empty set $R'$ and the set of candidate ASes $\mathcal{R}$ which satisfy the constraints listed in Section~\ref{ssec:routing} and are also contained in at least one $k$-connected subgraph of at least $N$ nodes, as only those can host one of the relay nodes of a $k$-connected network of $N$ relays. It then iteratively adds relays to $R'$ to maximize the number of covered attack scenarios while preserving $k$-connectivity for $R'$. This simple procedure runs in $\mathcal{O}(N)$ and works well in practice (Section~\ref{sec:net_eval}).

\remove{Our positioning algorithm works as follows. We first create a graph $G'(\mathcal{AS},E')$ in
which vertices corresponds to ASes with no customers and edges are direct peer-to-peer AS-links
Among those vertices, we consider as candidates only those that
are contained in at least one $k$-connected subgraph of at least $N$ nodes. Indeed, only those ASes can host one of the relay nodes of a
$k$-connected network of $N$ relays. 
 Next, we use a heuristic that iteratively adds relays to the set $R'$ to maximize number of attack scenarios and preserve intra-relay connectivity. 
As problem maximizing the attack scenarios maps to weighted maximum coverage problem, we use the well-known approximation of the latter and extend it to meet the connectivity-requirements by considering a subset of the candidates in each round. 

runs in linearly 
(with respect to $n$, the total number of relays) and works very well in practice as shown in ~\ref{sec:net_eval}.
}

In particular, in each round, we first select as candidates $\mathcal{R}'_k\subseteq\mathcal{R}_k\setminus R'$ which are connected with at least $\min\{k,\lvert R' \rvert\}$ of the already-selected ASes in $R'$. 
Then we add the candidate $r\in\mathcal{R}'_k$ that offers the maximum weighted extra coverage to $R'_{new} = R' \cup \{r\}$, i.e., the one with the maximum $\mathcal{C}\big(\bigcup_{r_i \in R_{new}'}\mathcal{S}(r_i)\big) - \mathcal{C}\big(\bigcup_{r_i \in R'}\mathcal{S}(r_i)\big)$ and we update the 
$R' := R'_{new}$. When we have selected all candidates, so that $\lvert R' \rvert = N$, we return $R'$.

\remove{
Our heuristic runs in line (with respect to $n$, the total number
of relays) and provides close-to-optimal results. }

We show in
Section~\ref{sec:net_eval}, that the resulting relay networks can readily protect
between 80\% to 98\%  of the existing Bitcoin clients (depending on the internal connectivity and number of deployed nodes) from 99\% of the potential
attackers. The complete pseudocode can be found in the Appendix~\ref{app:algos}, Algo.~\ref{algo:algo1}.

While our algorithm's goal is to minimize the attack vector rather than maximizing the deployment incentives, we discuss those in Section~\ref{sec:deployment}.

\begin{figure}[t]
 \centering
 \includegraphics[width=.5\columnwidth]{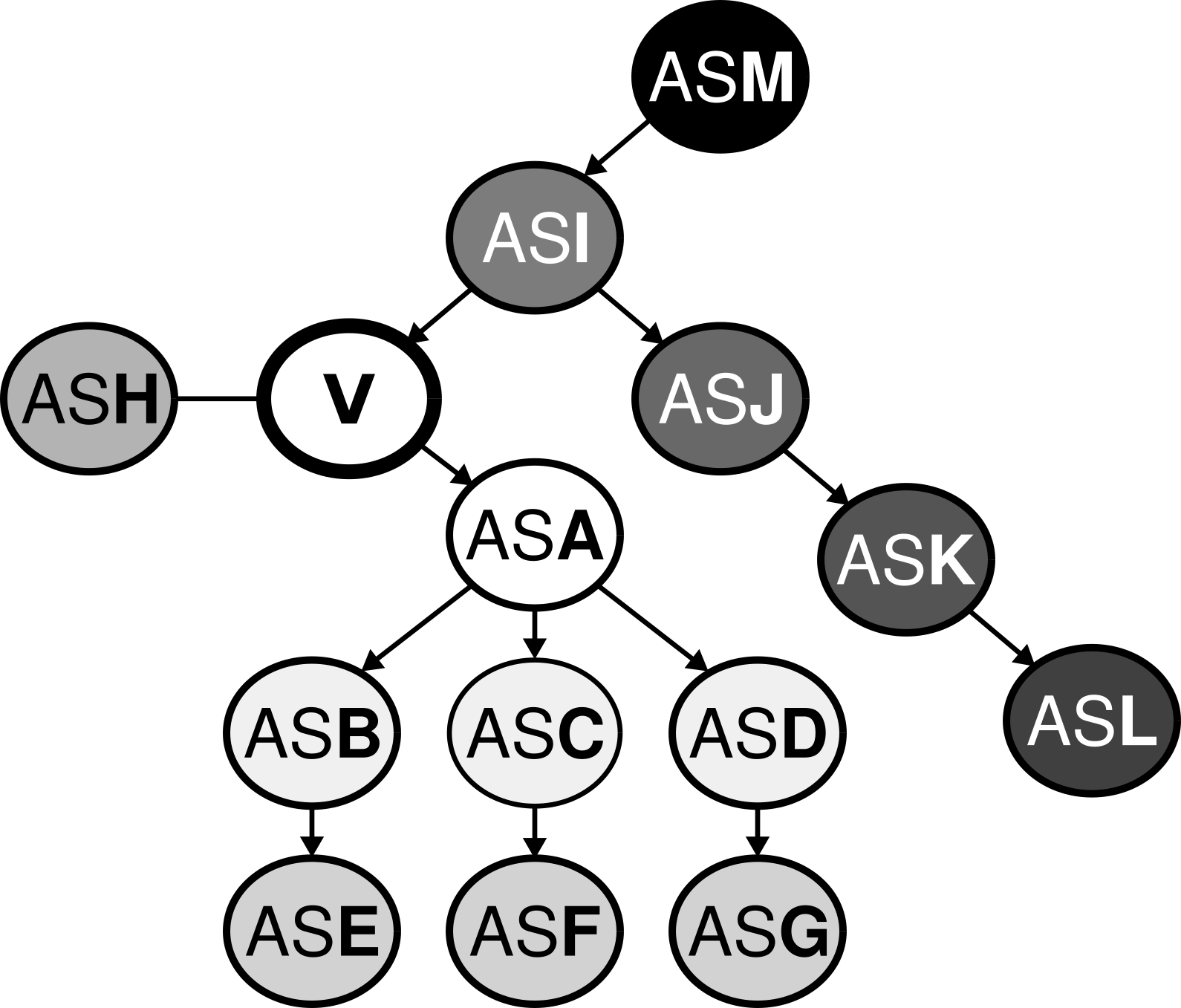}
 \caption{Shades illustrate preference. Traffic from $AS$ $V$ to a whiter AS is less likely to be hijacked}
 \label{fig:preferemce}
\end{figure}

\subsection{Calculating covered attack scenarios} 
\label{subsec:algo2}

Having explained how we can position \system relays based on the attack
scenarios they cover, we now describe how we compute these scenarios for each
relay, i.e., how we implement the function $\mathcal{S}$. Specifically, we now
describe how we compute the set of AS-level adversaries which can successfully
hijack traffic sourced from an AS hosting Bitcoin clients (say $AS$ $V$) and
destined to a relay AS (say $ASR$).

Our algorithm is based on the observation that if the fake advertisement
reaches the victim ($AS$ $V$), it will do so via the same propagation path as any
other prefix advertised by the attacker $ASM$ and will thus share its preference characteristics.
Therefore, to check whether the traffic from $AS$ $V$
is vulnerable, we only need to compare the path from $AS$ $V$ to $ASR$ with
the path from $AS$ $V$ to $ASM$. If the path to $ASM$ is more preferred 
by the last AS that the two paths have in common, then $ASM$ can successfully hijack traffic from an $AS$ $V$ to $ASR$.
This is because the last AS decides which 
of the two routes to use and advertise further. 
Observe, that the last ancestor is $AS$ $V$ itself if the paths from $AS$ $V$ to $ASR$ and to $ASM$ are disjoint.
The preference comparison is based on business relations among the on-path 
ASes and on the path length; namely, customers are preferred over peers; peers over providers; and shorter paths over longer ones.
\remove{
and on the path length which are the two factors that define the preference
for regular routing. 

In order to compare two paths we first find their last
common ancestor (possibly $AS$ $V$ or $ASR$ itself) which picks its preferred advertisement 
based on the above criteria, namely, customers are preferred over peers; 
peers over providers and shorter paths over longer.
}

As an example, Figure~\ref{fig:preferemce} illustrates an AS-topology which is augmented with the economic agreements between ASes. Arrows are reversed with respect to the money flow (if any) with providers being at the top and customers at the bottom. The different shades illustrate how preferred advertisements
originating from certain ASes are compared to others in the eyes of $AS$ $V$.
White is the most preferred and black the least. For example, $AS$ $V$ would prefer an
advertisement from its customer $ASA$ over any other advertisement ($ASA$ is
white). Thus, if a relay is hosted in $ASA$, no AS can divert the connection
to it. Among different originators that are reachable via a customer, the
shortest path will be preferred. As such, the first layer of customers is
lighter than the second, meaning that if a relay is hosted in, say, $ASE$, all
of $ASA$, $ASB$, $ASC$, $ASD$ and possibly $ASF$ and $ASG$ are effective possible
attackers. Similarly, if a relay is hosted in V's peer, namely $ASH$, all the
ASes in the customer cone of $ASA$ are possible attackers, as are ASes
of shorter path length in the peer cone. Finally, if the relay is reached via a
provider, length is not always relevant: In our example, $ASM$'s advertisement
will be less preferred than those from $ASA$-$ASH$, and even less
preferred than $ASL$'s one. Although both paths are via a
provider, namely $ASI$, and the path to $ASM$ is shorter, the victim will not use it. This 
is because $ASI$, which is the last common AS of the two paths,
prefers the path via $ASJ$ and will not advertise $ASM$.
The complete pseudocode can be found in the Appendix~\ref{sec:appendix}, Algo.~\ref{algo:algo2}.

\section{\system Resilient Relay Node Design}
\label{sec:node_design}

While a sophisticated software-based implementation  of the relay node might work\footnote{we discuss the possibility of a software deployment of \system in \S\ref{sec:deployment}}, it will have 2-3 orders of magnitude lower throughput ~\cite{netchain} compared to a hardware-based approach making it especially vulnerable to DDoS attacks. On the contrary, a hybrid implementation which utilizes programmable network devices can scale to billions of packets per second and mitigate malicious client directly in the data-plane, namely before they can reach the software component.

In this section, we explain the software/hardware co-design behind a \system relay node (Section~\ref{subsec:hw_sw_design}) and its operations (Section~\ref{subsec:operations}). While the protocol itself is specific to Bitcoin, similar techniques can be applied in other systems. We further discuss this issue in Section~\ref{sec:discussion}.

\begin{figure}[t]
 \centering
 \includegraphics[width=\columnwidth]{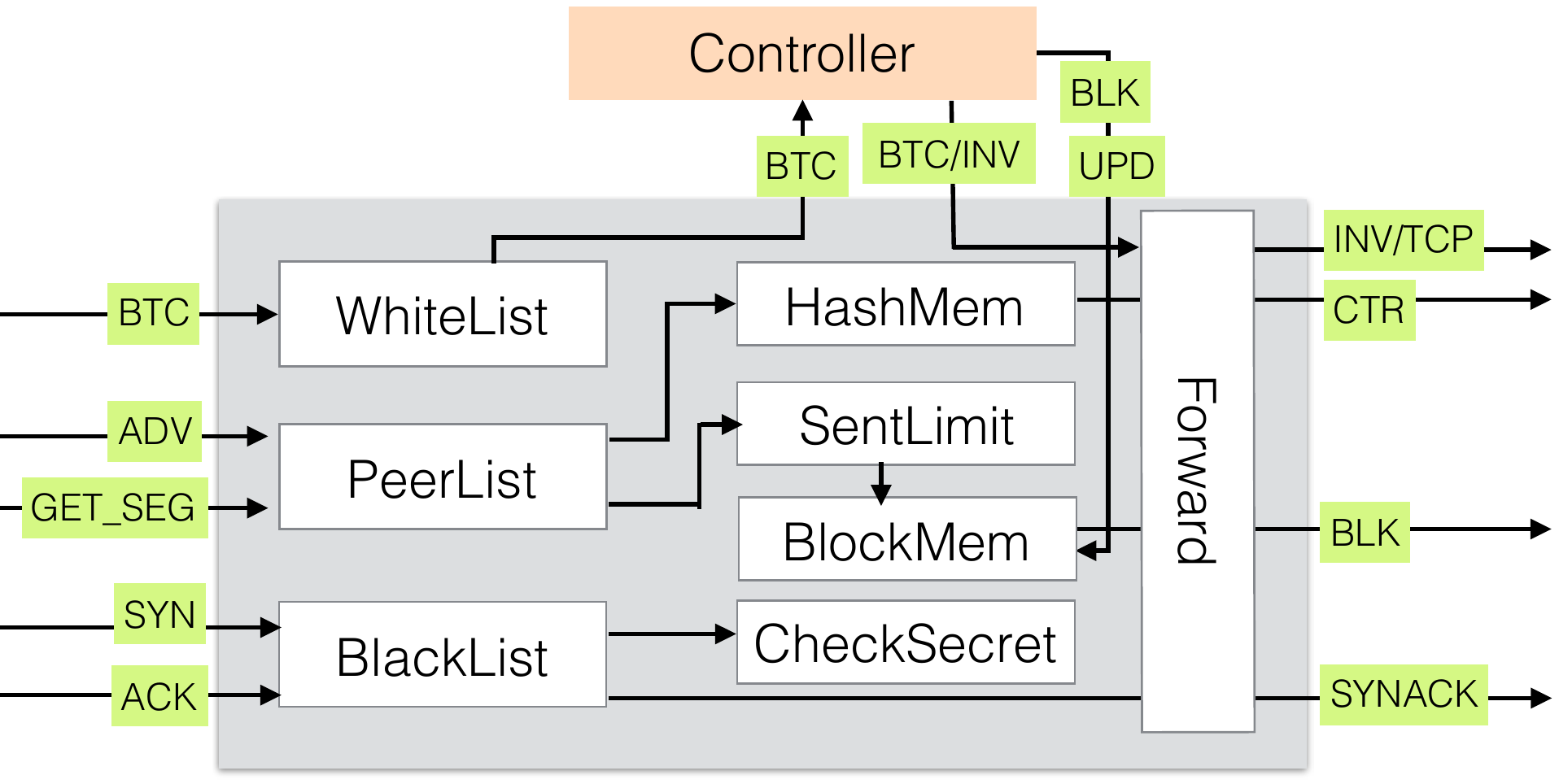}
 \caption{The switch intercepts all incoming traffic, answers to all UDP requests and redirects TCP traffic of whitelisted clients to the controller. The switch contains the latest mined Block in \emph{BlockMem} and multiple components to track the connected and banned clients (e.g.White/Black List, Connected) and detect attacks (e.g. CheckSecret, SentLimit)}
 \label{fig:nodeDesign}
\end{figure}

\subsection{Hardware/Software Co-Design}
\label{subsec:hw_sw_design}

Figure~\ref{fig:nodeDesign} illustrates \system's software/hardware co-design. It
is composed of a programmable switch connected to a
modified Bitcoin client, the Controller. 

The switch is responsible for: \emph{(i)} serving client connections;  \emph{(ii)}
protecting the controller from malicious clients; \emph{(iii)} propagating blocks; and \emph{(iv)} distinguishing new blocks from old ones. 
In contrast, the controller is responsible for 
validating new blocks, advertising them to the connected clients 
and updating the switch memory accordingly.

Relay clients establish UDP connections with the switch and rarely regular 
Bitcoin connections (over TCP) with the controller.
Switches only allow approved Bitcoin clients to establish connections with the
controller. As most clients ``consume'' blocks rather than producing them, we
expect most clients to only interact with \system through UDP connections.

\system's UDP-based protocol is composed of 8 messages: \textsf{SYN}, \textsf{SYN/ACK},
\textsf{ACK}, \textsf{NCONN}, \textsf{GET\_SEQ}, \textsf{ADV}, \textsf{UPD} and \textsf{BLK}. Similarly to TCP,
\textsf{SYN}, \textsf{SYN/ACK}, \textsf{ACK} are used to prevent spoofing attacks. \textsf{NCONN} is used for notifying the controller of new connections.
\textsf{GET\_SEQ}, \textsf{BLK} and \textsf{ADV} relate to block management. Specifically,
\textsf{GET\_SEQ} enables a client to request a particular segment of a block which is sent as a \textsf{BLK}, while
 \textsf{ADV} enables a client to advertise a newly mined block to the relay. Finally, a \textsf{BLK} message is also used by the controller after an \textsf{UPD} to update the switch with the latest block.

The switch maintains three data structures to manage client connections and
track down anomalies: \emph{PeerList}, \emph{Whitelist}, \emph{Blacklist}. \emph{PeerList} contains information about connected clients, i.e., those who successfully established the three-way handshake. In contrast, \emph{Whitelist} maintains information of clients that are allowed to communicate with the controller directly, while \emph{Blacklist} contains clients that have misused the relay and are banned. 
The switch also maintains one data structure to store the latest block(s): \emph{BlockMem}. \emph{BlockMem} is stored in indexed segments of equal size together with a precomputed checksum for each segment to allow the switch to timely reply with the requested segment avoiding additional computations. 
Moreover, the switch contains two components devoted to anomaly detection: \emph{SentLimit} and \emph{CheckSecret}.
\emph{SentLimit}, detects clients that requested a block too many times, while \emph{CheckSecret} calculates a hash for verifying whether the clients is using its true IP to connect to the relay node, during the handshake.
Finally, the switch also maintains one data structure for checking whether an advertised hash is known: \emph{Memhash}.

In the following, we describe the different operations performed by the relay
and how each of them modifies each of the data structures. In
Section~\ref{sec:node_eval}, we show that our design can sustain 1M malicious and 100k benign client connections with less than 5 MB of memory in the switch. This memory footprint is only a fraction of the memory offered by programmable switches today (tens of megabytes~\cite{jin2017netcache}), allowing the switch to implement other applications as well. 

\begin{figure}[t]
 \centering
 \begin{subfigure}[t]{0.45\columnwidth}
 \includegraphics[height=6cm]{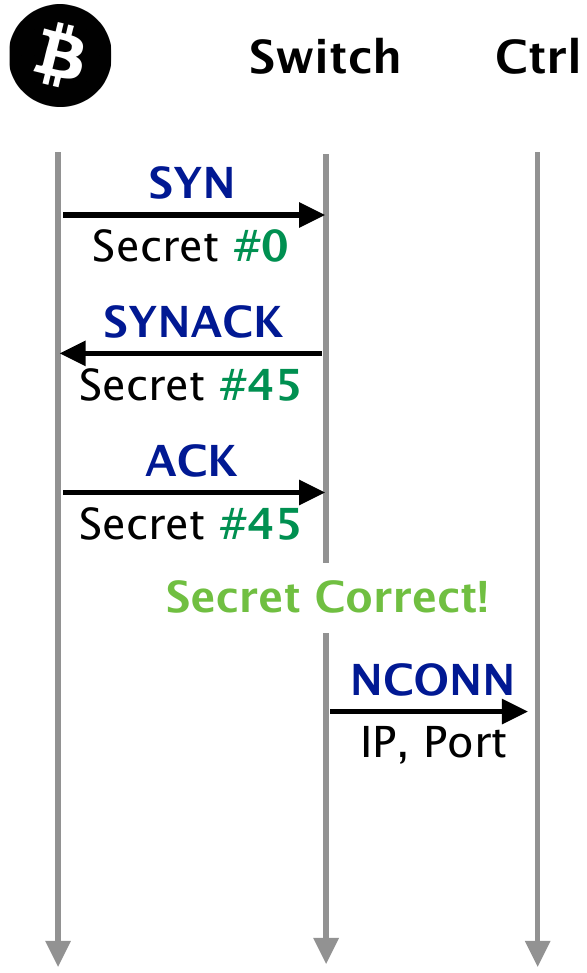}
\caption{Establishing connection}
 \label{fig:syn}
 \end{subfigure}
 \qquad
 \begin{subfigure}[t]{0.45\columnwidth}
 \includegraphics[height=6cm]{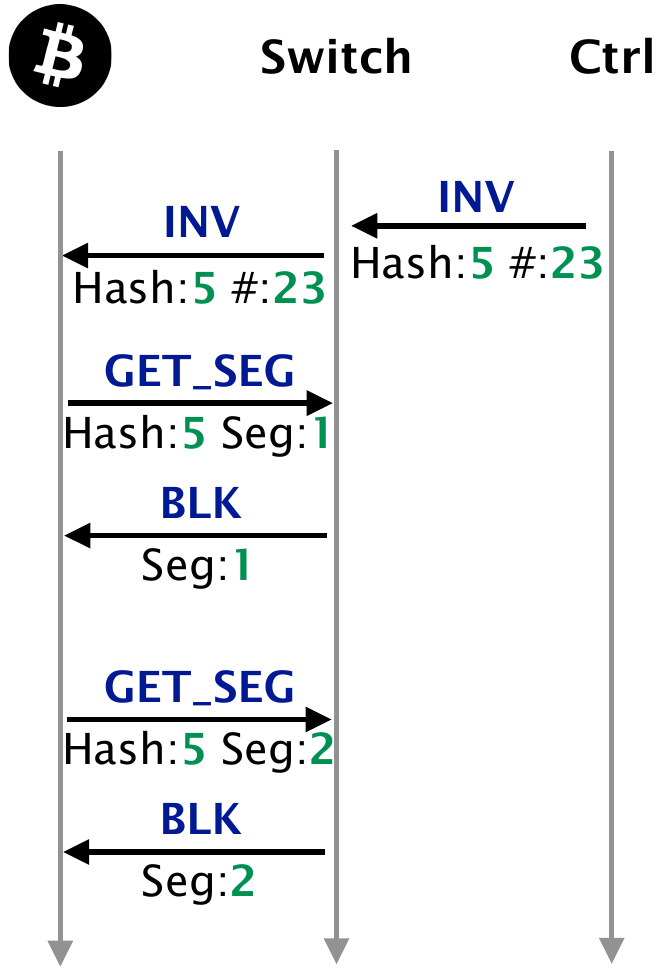}

 \caption{Transmitting Block}
 \label{fig:blk}
 \end{subfigure}
 \caption{(a) BTC client establishes a connection with the switch using a 3-way handshake. (b) Relay advertises a new block \textsf{INV} via the switch and transmits it using multiple \textsf{BLK} messages after client requests using \textsf{GET\_SEG} messages.}
\end{figure}

\begin{figure}[t]
 \centering
 \begin{subfigure}[t]{0.42\columnwidth}
 \includegraphics[height=6cm]{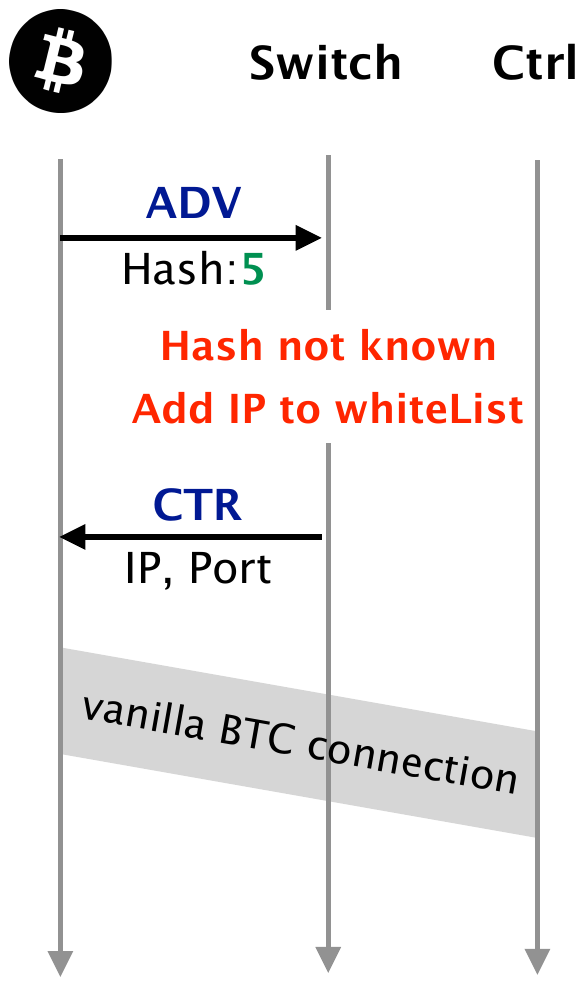}
  \caption{Receiving Block}
 \label{fig:blk2}
 \end{subfigure}
 \qquad
 \begin{subfigure}[t]{0.42\columnwidth}
 \includegraphics[height=6cm]{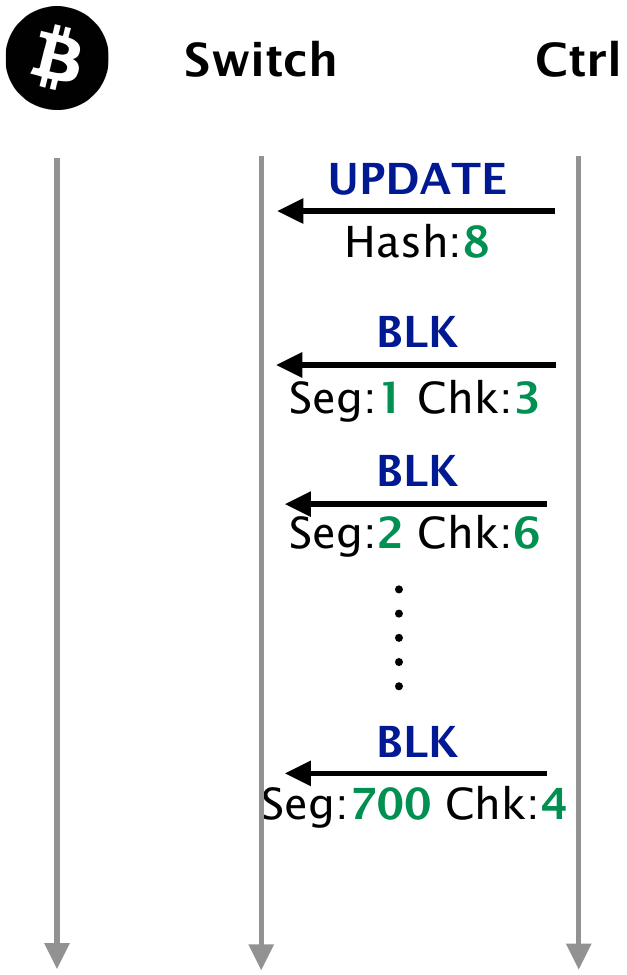}
 \caption{Updating switch}
\label{fig:upd}
 \end{subfigure}
 \caption{BTC client advertises a new Block. The switch identifies it as is an unknown block. The client gets white-listed and is thus permitted to connect directly to the controller as if it were a regular BTC client. If the received Block is valid: (b) the controller updates the switch using an \textsf{UPDATE} message carrying the Block's hash followed by a \textsf{BLK} messages carrying the data.}
\end{figure}

\subsection{Relay operations}
\label{subsec:operations}

We now describe \system relay operations in detail. The client and controller
are extended versions of the default Bitcoin client and the switch is
implemented in P4~\cite{bosshart2014p4}. Our protocol defines four operations:
\textit{(i)} how regular Bitcoin clients connect to a relay node; \textit{(ii)}
how a relay node propagates blocks back to them; \textit{(iii)} how a relay
node receives and validates blocks transmitted by the clients; and
\textit{(iv)} how the controller updates the switch memory upon the reception
of a new valid block. For each operation, the switch ensures that the relay's resources are not maliciously exhausted.

\myitem{Managing client connections} In order to avoid spoofing attacks, Bitcoin clients initialize
connections to relay nodes using a three way-handshake as shown in Figure~\ref{fig:syn}. 
As for a normal TCP connection, the client first sends a \textsf{SYN} packet.
Upon receiving the \textsf{SYN}, the switch echoes back a secret value calculated using the client's IP address and UDP port in a \textsf{SYN/ACK} packet. The client then includes this secret
value in the final (\textsf{ACK}) packet as a proof that it owns the source IP address that it is using.

Upon successfully completing the handshake, the switch adds an entry for the
client's IP and port number in the \emph{PeerList} and notifies the controller via a \textsf{NCONN}
message. The \emph{PeerList} is implemented as Bloom filter
(BF) for memory efficiency. As such, it enables the switch to verify that an incoming packet belongs to an established connection and drop it otherwise. BFs do not support listing all inserted items. as such the controller needs to store the connections for future use
(e.g., advertising new blocks and updating the \emph{PeerList}).

\myitem{Learning new blocks} Relay nodes need to learn new blocks that are mined. New blocks are transmitted to the relays from regular clients. 
Being a network device with limited computational capabilities, the switch is unable to validate blocks. Thus, advertised blocks need to be transmitted to the controller after they have been filtered by the switch.

As illustrated in Fig.~\ref{fig:blk2}, the node advertises a block by its hash to the switch using an \textsf{ADV} message. The switch checks whether the hash is already known using the HashMem. If the hash is not known, then the switch asks the client to connect to the controller with a \textsf{CTR} message and stores its IP in the \emph{Whitelist}. If the transmitted block is legitimate the client's IP will stay in the whitelist for four days.
The client connects to the controller as if it was a regular Bitcoin client, while the switch forwards the TCP traffic to the controller. The switch only allows packets from white-listed clients to reach the controller.
Observe that a malicious miner cannot monopolize or overload the controller with its connections as even a pool with 30\% of the hash power cannot keep more than $172$ whitelisted clients in any given moment.\footnote{Every day, 144 Blocks are mined (on average). For each block at most one node is whitelisted (the one that is not already whitelisted and advertised the Block first)}

Yet, a malicious miner might still try to engineer block races by flooding the relay node with multiple blocks simultaneously which will need to be validated by the controller. To shield against this attack, the switch keeps the number of active nodes that are white-listed. When this number exceeds a predefined threshold set based on the controller's hardware capabilities, the switch will stop whitelisting new clients. In this case, the controller receives blocks from the nodes that are already whitelisted. These nodes should be diverse enough, with respect to mining power origin, to keep the relay up-to-date, thanks to the expiry mechanism in the \emph{Whitelist}. For instance, any pool with at least 0.17\% of mining power can keep at least one node in the \emph{Whitelist} forever. In essence, the switch implements a simple yet efficient reputation-based access-list to protect the controller from Sybil attacks.

\myitem{Updating switch with a new block} If a newly-transmitted block is
valid, the controller updates the switch's memory with a new mapping between
segment ID and block segment data that corresponds to a particular block hash.
The switch can then transmit the segments to the clients upon requests. Observe
though that the switch sends data to a UDP socket. Thus, the IP and UDP checksums
need be correct for the packet to be accepted. The UDP checksum is
calculated using a pseudo-header and the one's complement sum of the payload
split into 16 bits segments. Because computing this in the switch would result in
too many computations, we cache the precomputed the one's complement sum of the block
segment together with the segment itself. Using this value the switch needs
only to add the header parts that are different per client.

Figure~\ref{fig:upd} illustrates the sequence of packets the controller sends to update the switch. Initially, it sends an \textsf{UPDATE} message containing the new hash. This first message tells the switch to prepare its state for the new block. The next messages are sent to transmit each of the segments of the block as well as a precomputed one's sum of it.

\myitem{Propagating a newly-learned block} The relay node advertises new blocks to all its connected clients who can then request a block segment-by-segment.
Blocks are transmitted in multiple individual segments for three reasons: \emph{(i)} to allow clients to request lost segments independently; \emph{(ii)} to avoid loops in the data plane which would be otherwise needed as the block does not fit in one packet; and \emph{(ii)} to protect against amplification attacks.

As illustrated in Figure~\ref{fig:blk}, the controller sends an \textsf{INV} message which is forwarded by the switch. This \textsf{INV} message contains the hash of the new block as well as the number of segments necessary. In the example, the relay advertises hash $\#5$ which is composed of $23$ segments. If the Bitcoin client is unaware of the advertised block, it requests it using a \textsf{GET\_SEG} message containing the hash of the block and each of the $23$ segment IDs. In the example, the client first requests the segment of $ID$$:$$1$ of the block with hash $\#5$ then the segment of $ID$$:$$2$ and so on. If either the \textsf{GET\_SEG} or the \textsf{SEG} is lost the client will simply request the corresponding segment again.

As a protection mechanism, the switch bans clients that request a block multiple times. To that end, all requests traverse a heavy-hitter detector, namely \emph{SentLimit}. In \system, we just reuse a component optimized for programmable switches~\cite{hit} which can operate with just 80KB of memory.

\section{Network Architecture Evaluation}
\label{sec:net_eval}

In this section, we evaluate \system's efficiency in protecting Bitcoin against
routing attacks. Specifically, we answer the following questions: How
effective is \system in preventing routing attacks targeted against the
entire network and individual clients? How does this effectiveness change with
the size and the connectivity of the \system network? How does \system stand out
against other relay networks and known counter-measures?

We found that even a small deployment of $6$ single-connected \system nodes can
prevent 94\% of ASes in the Internet from isolating more than 10\% of the Bitcoin
clients; while larger deployments of $30$ relays that are 5-connected can
prevent more than $99\%$ of the ASes from isolating more than 20\% of Bitcoin
clients. In addition, we show that existing relay networks, like
Falcon~\cite{falcon} and FIBRE~\cite{fibre}, offer \emph{no} protection against
routing attacks. Finally, we show that \system provides security level on-par
with hosting all clients in /24, an effective but clearly impractical solution.

We start the section by describing our methodology (Section~\ref{ssec:methodology}) before presenting our results in detail.

\subsection{Methodology}
\label{ssec:methodology}

\myitem{Datasets} Our evaluation relies on a joint dataset combining routing
and bitcoin information. 
Regarding routing information, we rely on the AS-level topology and AS-level
policies provided by CAIDA~\cite{caida_as_level}, collected in May 2018.
We rely on the routing tree algorithm~\cite{goldberg_how_secure_are_interdomain_routing_protocols} to compute the
forwarding path followed between any two ASes. We consider that
the paths originated by an attacker AS are systematically picked at the
tie-breaking state of the BGP decision process (the worst-case for \system)
\footnote{Results for the opposite case, where tie-breaking systematically picks paths originated by relay ASes, can be found in Appendix \ref{app:graphs}}.
Regarding Bitcoin information, we rely on the IPs of Bitcoin clients
from~\cite{btc_site} along with the IPs of existing relay nodes from~\cite{fibre,falcon}, both collected in May 2018.

We merge the two datasets by associating each Bitcoin IP to the AS advertising
the most-specific IP prefix covering it (using the routes collected by RIPE BGP collectors~\cite{ripe:ris}).

\begin{figure}[t]
 \centering
 \includegraphics[width=\columnwidth]{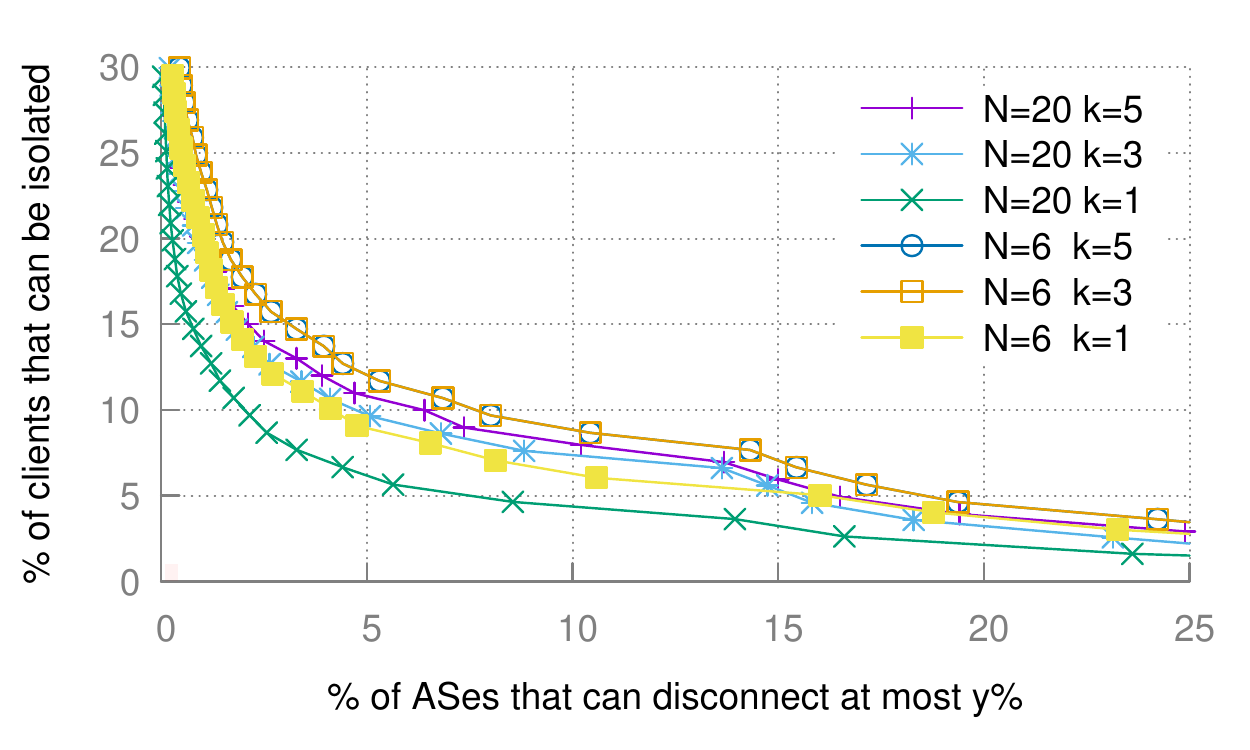}
	 \caption{Less than 2.5\% of ASes are able to disconnect more than 15\% of clients. Thus, the chance that a random AS-level adversary will be able to simultaneously hijack many clients is extremely low. ($N$: the number of deployed relays; $k$: relay-graph connectivity; Tie breaks in favor of the attacker)}
	 \label{fig:relays:aof1}
\end{figure}

\subsection{\system security efficiency}
\label{ssec:security_efficiency}

\myitem{\system protects against network-wide partitions} To evaluate how effective \system is against adversaries that wish to partition
the Bitcoin network, we quantify how likely it is for a random adversary to be able to
disconnect multiple clients from the relay network. The fraction of clients a
particular AS can disconnect from the relays poses an upper bound to the maximum partition that she can create in the Bitcoin network, as Bitcoin nodes connected to the relay network cannot be partitioned.

Fig.~\ref{fig:relays:aof1} illustrates how protected the Bitcoin network is
depending on the size $N$ and internal connectivity $k$ of the \system network. 
The graph shows, for each given fraction y of Bitcoin nodes, what percentage of ASes would be able to independently disconnect it from \system .

For $N=20,k=1$, less than 3\% of ASes are able to prevent a considerable
fraction of Bitcoin clients (15\%) from connecting to the relay network. In
contrast, more than $90\%$ of the clients can be isolated by \emph{any} AS in
the current network~\cite{hijackbtc2017}. 

The mapping between the number of possible attackers and the partition sizes
varies with the size and connectivity of \system. In particular, increasing the
number of deployed nodes decreases the chances that adversaries can divert
traffic successfully. On the other hand, decreasing the intra-connectivity requirements
(i.e., the value of $k$) allows our algorithm (Section~\ref{sec:routing}) to select
from a larger set of relays and thus to form a more effective \system. This
creates an interesting trade-off between how secure the intra-relay
connectivity is and how well the relays cover the existing Bitcoin network. For
example, while a \system of $6$ relays that are connected in full-mesh
(5-connected graph) is extremely hard to partition, as the AS-level adversary
would need to divert $5$ peer-to-peer links, it enables more AS-level
adversaries to disconnect a large part of Bitcoin clients from \system. For
example, 3\% of ASes can potentially create a partition including 22\% of Bitcoin nodes. In contrast, a
1-connected \system allows fewer attackers to perform severe attacks---only 1\% of ASes could create a 12\% partition---but can be partitioned by a
single link failure or successful hijack from a direct peer.

\begin{figure}[t]
 \centering
 \includegraphics[width=\columnwidth]{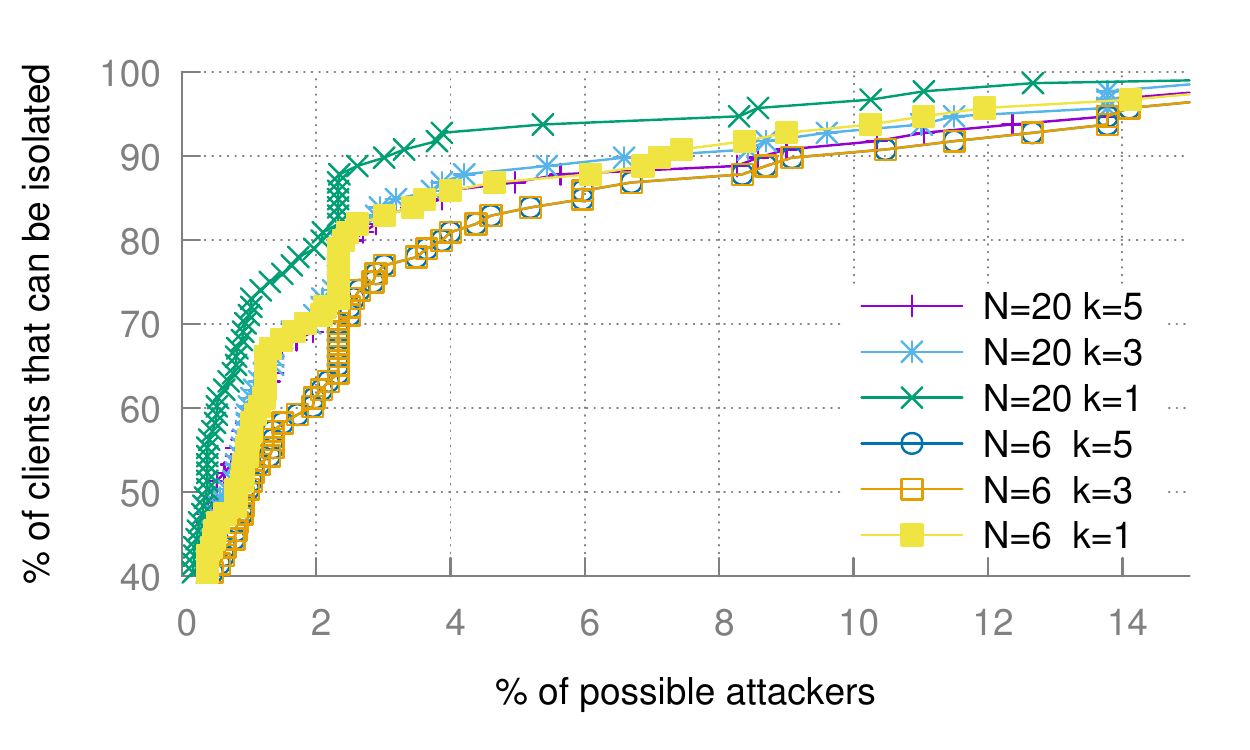}
	 \caption{85\% of the clients are protected against 96\% of possible attackers (Tie breaks in favor of the attacker)}
	 \label{fig:relays:vsf1}
\end{figure}

\myitem{\system protects most individual clients} To evaluate how effective \system
protects individual clients, we look at how likely it is for Bitcoin clients to be prevented by a random AS-level adversary from reaching \emph{all} relay nodes. 

\begin{figure}[t]
 \centering
 \includegraphics[width=\columnwidth]{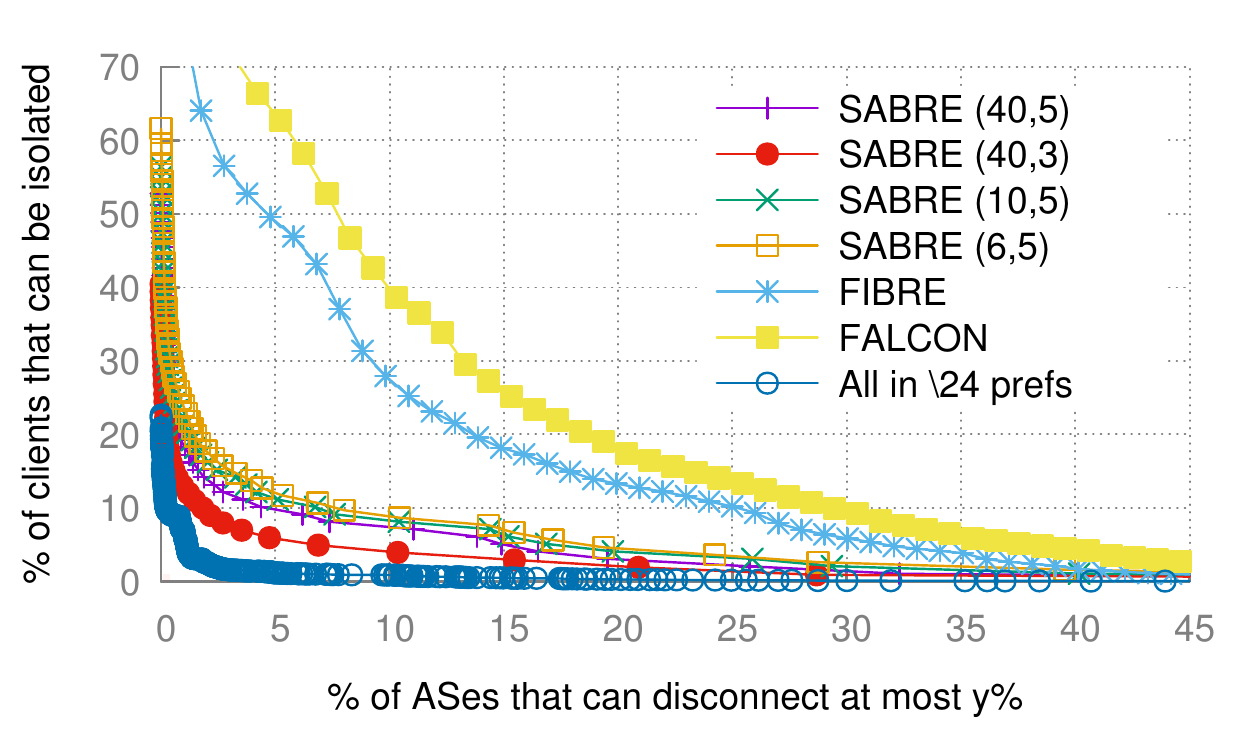}
	 \caption{\system is far more secure than deployed relays and very close to the unemployable alternative countermeasure of hosting all clients in $/24$. (Tie breaks in favor of the attacker)}
	 \label{fig:relays:aof2}
\end{figure}

Fig.~\ref{fig:relays:aof2} shows, for a given percentage of ASes, what percentage of Bitcoin clients could be attacked and disconnected from \system by this percentage of ASes. 

We see that 80\% of the clients are protected from 96\%
of the AS-level adversaries even with a \system network of only $6$ nodes that
are 5-connected. There is again a trade-off between secure intra-connectivity
and the coverage of Bitcoin clients. For example, a \system of $6$
1-connected nodes protects 90\% of Bitcoin clients from 92.5\% of ASes,
while a fully connected 6-node \system protects from only 89.5\% of ASes.
Interestingly, increasing connectivity from $k=3$ to $k=5$ does not
decrease the protected clients significantly while making disconnecting the
relay network almost impossible.

\begin{figure}[t]
 \centering
 \includegraphics[width=\columnwidth]{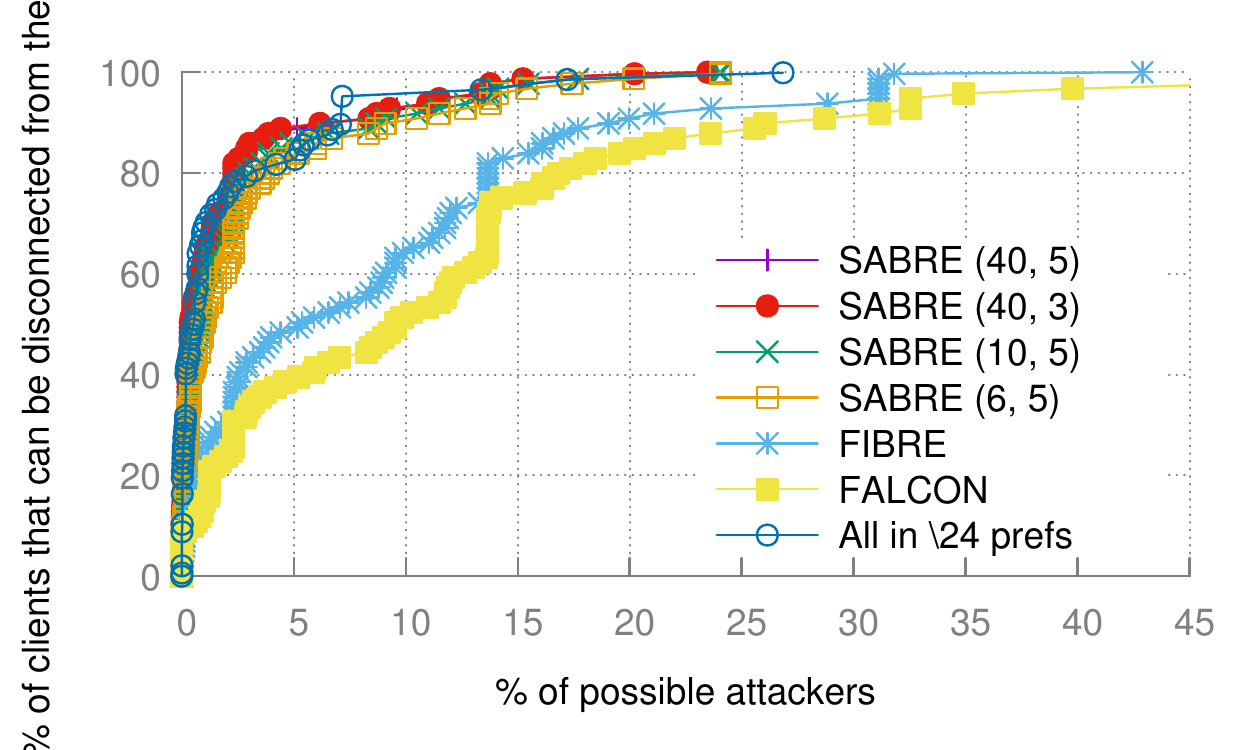}
	 \caption{Falcon does not protect many clients as it is centralized to only two ASes. \system  performs on-par with hosting all clients in $/24$ while being deployable (Tie breaks in favor of the attacker)}
	 \label{fig:relays:vsf2}
\end{figure}

\subsection{\system efficiency compared to existing relay networks}
\label{subsec:net_eval_others}
We compare \system to FIBRE~\cite{fibre} and Falcon~\cite{falcon} with respect to their effectiveness against routing attacks. We found that \system outperforms both, for three key reasons.

\myitem{Existing relays are vulnerable to longer-prefix hijacks} \emph{All}
relay nodes of both FIBRE and Falcon are hosted in prefixes that are shorter
than $/24$. As such, \emph{any} AS-level adversary can partition the relay
nodes from each other as well as from the Bitcoin clients only by hijacking $6$ more-specific
prefixes for FIBRE and $10$ for Falcon.

\myitem{Existing relay networks are poorly connected} Even if these relay networks were to host their nodes in /24 prefixes, we found out that their
connections could still be diverted by same-prefix advertisements. In
particular, we found that FIBRE relays would be disconnected by any of $652$
ASes, and Falcon by any of $3$ ASes even if $/24$ prefixes were used. 

\myitem{Existing relays provide bad coverage} Again assuming that these relay
networks would host their nodes in $/24$ prefixes, their client-to-relay
connections would still be more vulnerable than those of \system allowing for
more network-wide and targeted attacks. We compare those relay networks with
\system with respect to how well they protect against routing attacks using the
same graphs as in Section~\ref{ssec:security_efficiency}. In particular, Fig.~\ref{fig:relays:aof2} shows the percentage of ASes that are able to independently isolate a fraction of the Bitcoin network as a function of this fraction while Fig.~\ref{fig:relays:vsf2} shows the cumulative percentage of clients as a function of the number of AS-level adversaries that could disconnect them from all relay nodes. While
FIBRE is slightly better than Falcon, \system outperforms both.

\subsection{\system efficiency compared to hosting all clients in /24s} We now
compare \system to the most effective countermeasure against routing
attacks: hosting \emph{all} bitcoin clients in /24
prefixes~\cite{hijackbtc2017}. While effective, this countermeasure is also
highly impractical as it requires ISP cooperation in addition to increasing the
size of the routing tables Internet-wide. We found that \system offers comparable level of protections against network-wide and targeted attacks while being easily deployable. 

The comparison between the two approaches is not straightforward as \system protects the network even if the attacker has already partitioned the Bitcoin Peer-to-Peer network while the other approach aims at securing the Peer-to-Peer network itself. In the following, we describe our methodology and key results.

To compare the two alternatives with respect to their protection against partition attacks we need first to identify the AS-level adversaries that would be able to isolate a considerable fraction of Bitcoin clients using same-prefix advertisements only. To do so, we use a breadth-first search on the AS-level topology graph, which traverses the graph in order of descending preference (Section~\ref{subsec:algo2}). We run the traversal from every AS with Bitcoin clients $X$. All ASes that are traversed by the search before another AS with Bitcoin clients are able to isolate $X$ from the Bitcoin network. This calculation gives only a lower bound with respect to the possible partitions, i.e. hosting all clients in $/24$ prefixes might offer less security than what we computed. Our results are included in Fig.~\ref{fig:relays:aof2}. Indeed, hosting all clients in $/24$ prefixes would secure the Bitcoin Network better than \system, as partitions larger than 20\% would be possible for only 0.016\% of ASes. 

In order to compare how many attackers can successfully isolate individual Bitcoin clients, we looked at the ASes that are able to divert traffic from each of those clients to all others in the network. The results are included in Fig.~\ref{fig:relays:vsf2}. The two approaches show similar protection levels with \system being slightly better at times. This is because \system can place relays in any AS in the Internet, while the alternative countermeasure is limited to the actual distribution of Bitcoin clients.

\section{Software/Hardware Co-design Feasibility}
\label{sec:node_eval}

We validated the feasibility of our co-design by testing it in practice using regular and modified  Bitcoin clients connected to the \system components, namely the P4 switch and controller.
We showcase that \emph{(i)} a programmable switch can seamlessly talk to a Bitcoin client without any software interaction; and that \emph{(ii)} the data-plane memory footprint is low compared to the on-chip memory available in today's programmable switches. 

\begin{figure}[t]
 \centering
 \includegraphics[width=.9\columnwidth]{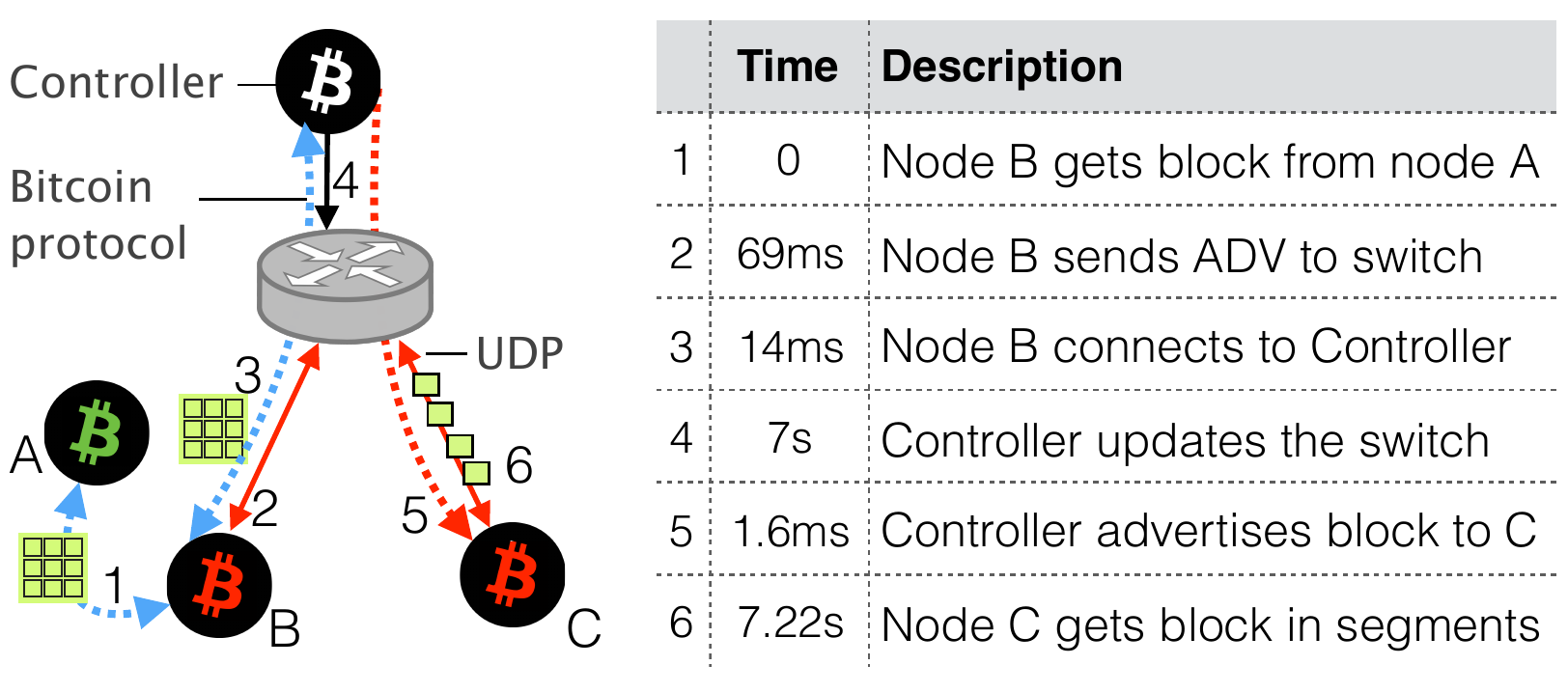}
	 \caption{A block can be successfully transmitted from node $A$ to node $C$ via the \system after it has been validated by the controller.}
	 \label{fig:testbed}
\end{figure}

\myitem{Implementation/Testbed} Both the controller and the clients are implemented as extensions of the default Bitcoin client version 0.16. The former containing $\approx$650 added or modified lines of C++ code and the latter $\approx$680 lines. The switch is implemented in $\approx$900 lines of P4 code.
Our prototype runs on Mininet~\cite{lantz2010network} and
uses the publicly available P4 behavioral model (\textsf{BMV2})~\cite{p4lang}
to emulate the switch. Our testbed (see Fig.~\ref{fig:testbed}) is composed of three clients $A$, $B$, $C$ along with a relay-node consisting of a switch and a controller. Nodes $B$, $C$ (shown in red) are modified and are connected to \system, while node $A$ (shown in green) is an unmodified Bitcoin client. The controller is also a modified Bitcoin client  

\myitem{Timing} We walk through the life of a block that was mined in the
Bitcoin network and sent by the unmodified client $A$, to a modified one, namely node $B$. The latter will advertise the new block to the
switch, which will allow node $B$ to connect directly to the controller and transmit it. The
controller will then update the memory of the switch and will advertise the
Block to the connected peers (e.g., C). Next, node $C$ will request and receive the
block in segments. The main steps of this procedure are listed in
Fig.\ref{fig:testbed} which contains for each step an index, a description and the time spent for it in our prototype implementation. The most time-consuming
operations are updating of the switch and transmitting of the Block,
taking $\approx$$7$s each. These relatively large times
are because we rely on a simulated version of a P4 switch. The only actual bottleneck in a hardware implementation would be the uplink of the relay nodes.

\myitem{Memory requirements}
We analytically calculated the memory for each of the components in the switch taking into consideration the expected load. Table~\ref{tab:mem} summarizes our results. It comprises the name of the component, its capacity, i.e., the number of elements that can be added such that the false positive rate listed in the third column is not exceeded. We can see that the cumulative memory needed does not exceed 5MB which is within the limitations of today's programmable switches.
The most memory-demanding component is the \emph{Blacklist} for which we budget 1 million entries. This is necessary to allow for mitigating DDoS attacks. In contrast, the components devoted to regular operations are less memory-demanding since the number of legitimate clients is significantly less. For instance, we only reserve space for $100k$ clients in the \emph{PeerList} and $100$ for the \emph{Whitelist}; both require less than 1MB. Observe that Bloom filters for regular clients have a lower false positive rate than the \emph{Blacklist}. This enables to serve already connected clients even if the switch is under such an extreme DDoS attack that the \emph{Blacklist} is flooded. We do not list the requirement of the \emph{SentLimit} component as they are negligible~\cite{hit}. Finally, the memory needed for the storing the latest block as well as for keeping all known hashes takes about 1MB each.

\begin{table}[t]
\centering
\footnotesize
\def\arraystretch{1.1}
\setlength{\tabcolsep}{7.5pt}
\begin{tabular}{l l l l }
\toprule

\begin{tabular}{@{}l@{}} \emph{Component}  \end{tabular} & \begin{tabular}{@{}l@{}} \emph{Items}\end{tabular} &  \begin{tabular}{@{}l@{}} \emph{False Positive}  \end{tabular} & \begin{tabular}{@{}l@{}} \emph{Memory} \end{tabular}  \\
\midrule
BlackList: & 1000000 & 0.001 & 1.80MB\\
WhiteList: & 100 & 0.0001 & 239.75B\\
HashMem: & 518823 & 0.0001 &  1.24MB\\
PeerList: & 100000 & 0.0001 & 479.25K\\
BlockMem: & 1 & - & 1.0MB\\
\bottomrule
\end{tabular}
\caption{The memory used in the P4 switch is always \textless5MB}
\label{tab:mem}
\end{table}

\section{Deployability \& Incentives}
\label{sec:deployment}

Similarly to existing relay networks (e.g. Fibre~\cite{fibre},
Falcon~\cite{falcon}), \system requires one or more entities to deploy and maintain its relay nodes. We now discuss the incentives and the
practicality aspects underlying a \system deployment.

Given the amount of money at stake,\footnote{As of 7 August 2018 Bitcoin market
capitalization accounts for more than 120 billion of dollars.} and the
devastating effects of routing attacks~\cite{hijackbtc2017}, important Bitcoin
clients---particularly miners---have an incentive to finance the hosting costs
of \system. ASes have therefore an incentive to offer a \system service for a
fee. We argue that such ``business model'' is reasonable for two reasons.

First, we note that a large number of ASes that were found to be good
candidates for hosting relays are cloud providers, CDNs, IXPs, large ISPs, or
Software-as-a-Service (SaaS) providers. This should come as no surprise as such
ASes are actively trying to establish as many peering connections as possible
to improve their services. Deploying \system nodes in such ASes is practical as
they already sell online services or are research-friendly
(IXPs~\cite{gupta2015sdx}).
Moreover, even if some eligible ASes do not consent to host \system nodes, the
effectiveness of \system will not be affected as: \emph{(i)}
\system only requires few nodes to be useful (as little as $6$ ASes, see Section~\ref{sec:net_eval}); and \emph{(ii)} there are more than $2000$
possible locations for hosting ASes. In short, no candidate AS is irreplaceable.

Second, cloud providers already experiment with renting our advanced hardware
resources. As an illustration, Amazon EC2 offers the possibility to rent
hardware-accelerated computing instances with field-programmable gate
arrays~\cite{amazon_f1_instances}. We therefore envision that cloud providers
could also rent out hardware-accelerated computing instances with programmable
network data planes. If this is not the case, a pure software-based
implementation of \system would still protect the Bitcoin network from routing
attacks leaving DDoS to traditional solutions. Such software-based
implementation of \system can readily be deployed as it only requires the
possibility to host virtual machines.

Finally, we note that while \system requires some effort to be deployed, it is
much more practical than known alternative solutions~\cite{hijackbtc2017} such
as requiring all ASes to deploy secure routing protocols or expecting all ASes
with Bitcoin clients to filter their routes.
\section{Discussions}
\label{sec:discussion}

We now provide answers to high-level concerns about \system, including how much
of its design generalize to other Blockchain systems.

\myitem{Doesn't \system centralization violate the decentralization premises of
Bitcoin?} \emph{No}, for three main reasons. First, it acts alongside Bitcoin and does
not aspire to replace or compete with the existing peer-to-peer network.
Instead, \system enhances the connectivity and significantly reduces the attack
surface which would otherwise allow any AS-level adversary to partition the
network. Second, more than one \system-like systems can harmoniously co-exist,
each belonging to a different entity. Indeed, regular clients or mining pools
can connect to them at the cost of a lightweight UDP connection. Third, \system
has the potential to allow less well-connected miners to get their fair-share
out of the block-rewards making it less likely for others to engineer block
races.

Observe that existing relays (e.g. Fibre~\cite{fibre},
Falcon~\cite{falcon}) are small in size and controlled
by a single entity, just like \system. Neither of these characteristics
prevented them from having significant and positive impact to Bitcoin, namely
decreasing its orphan rate. 

\myitem{Why focusing on Bitcoin?} We focus on Bitcoin as opposed to more
state-of-the-art cryptocurrency (e.g., Ripple~\cite{ripple} or Ethereum~\cite{etherium}) for
three main reasons. First, Bitcoin is extensively studied~\cite{miller2015discovering}, ~\cite{decker2013information}, ~\cite{neudecker2016timing} and the
effectiveness of routing attacks against it is
well-understood~\cite{hijackbtc2017}. In contrast, more
sophisticated Blockchain systems (e.g., Bitcoin-NG~\cite{eyal2016bitcoin},
Ouroboros~\cite{kiayias2017ouroboros}, OmniLedger~\cite{kokoris2017omniledger},
Algorand~\cite{gilad2017algorand}) are not yet deployed at large scale and thus
their exact routing characteristics are unknown. Second, Bitcoin remains the most widely used cryptocurrency making its security paramount for many users.  

Still, one could argue that Bitcoin is a notoriously slow-moving community.
Yet, deploying \system does not need to be approved by the community as a
whole. Indeed, connecting to a \system node simply requires an extended client
establishing lightweight UDP connections.

\myitem{Can \system protect other blockchain-based systems from routing
attack?} \emph{Yes}. Although our work focuses on Bitcoin, many of \system
design principles are applicable to other Blockchain systems. Next, we
separately discuss the generality of the network and the node design.

\system network design is useful for any Blockchain system as it allows them to
mitigate partition attacks~\cite{hijackbtc2017}. Partition attacks are a threat
not only for permissionless or unencrypted blockchain systems like Bitcoin, but
also for permissioned and/or encrypted ones. In fact, the only Blockchain
systems that are not vulnerable to BGP hijacks are those whose nodes are all
hosted within a single AS or corporate network as their traffic is not routed
via BGP. Specifically, \system allows nodes to exchange information even if a
malicious AS-level adversary hijacks and drops traffic among them. In fact, the
properties upon which the \system network is built can also be used by miners
to interconnect and/or host their mining power, or by new Blockchain systems to
place their nodes in a routing aware manner that is inherently robust to BGP
hijacks. Finally, \system network design would be useful even to the most advanced Blockchain systems (e.g., ByzCoin~\cite{kogias2016enhancing}, OmniLedger~\cite{kokoris2017omniledger}) that can mitigate the effects of partition attacks by detecting the attack and freezing commits, effectively turning the attack into a Denial of Service. In particular, \system would allow them to retain liveness during the attack as resolving a BGP hijack is a human-driven process that can take hours~\cite{hijackbtc2017}.

In contrast to its network design, \system node's design is to some extent
specific to Bitcoin. For example, systems whose traffic is encrypted cannot be
served exclusively from a programmable network device. Even so, \system's node
design exhibits two key properties that all blockchain system share and onto
which further systems can be built. First, blockchain systems are
communication-heavy (due to the need to reach consensus) meaning that the use
of programmable switches can increase the throughput by offloading
communication burden to the hardware. Second, most popular items tend to be
predictable, as most nodes will always request the latest mined content, making \system-like caching strategies very effective.
\section{Related work}
\label{sec:location}
\remove{
\myitem{Relay Networks} Connected to relay networks (such as
Falcon~\cite{falcon} and FIBRE~\cite{fibre}) to obtain better performance
guarantees than the one provided by the vanilla Bitcoin network is a common
thing amongst Bitcoin clients. As an illustration, more than 36\% of the entire
hashpower in Bitcoin is already connected to the Falcon relay network so as to
speed up message propagation~\cite{}. Yet, to the best of our knowledge, we are
the first to show that relays can also improve the security properties of the
network, not only its raw performance.
}

\myitem{Using P4 switches as cache}
Previous works have used programmable network devices to cache values including Netcache~\cite{netcache} and NetChain~\cite{netchain}. Netcache use Tofino switches~\cite{tofino} as a cache for key-value stores, enabling them to deal with skewed requests in memcached applications. 
Similarly, NetChain~\cite{netchain} caches key-values stores in switches to
boost Paxos consensus protocols used in data centers to coordinate servers. In \system, we also rely on switches to cache information (here, blocks) which we complement with a novel UDP-based retrieval protocol and a dynamic access list.

\myitem{BGP security} Many proposals have been proposed over the years to reduce or prevent routing attacks. We distinguish two approaches: origin validation and path validation. Origin validation~\cite{rfc6480} relies on RPKI~\cite{rpki}, a X.509-based hierarchy mapping IP prefixes to ASes, to enable the routers to filter BGP advertisements originated from unauthorized ASes.
Path validation~\cite{bgpsec} secures BGP by adding cryptographic signatures to the BGP messages. It allows the recipient of an announcement to cryptographically validate that: (i) the origin AS was authorized to announce the IP prefix; \emph{and} (ii) that the list of ASes through which the announcement passed were indeed those which each of the intermediate AS intended.
Unfortunately, none of these proposals have been widely deployed, leaving the Internet still widely vulnerable to routing attacks~\cite{sharon}. In contrast, \system enables to secure Bitcoin against routing attacks today, without requiring all ASes to agree or change their practices .

\myitem{Routing attacks on ToR} Extensive work has been done on routing attacks on 
ToR~\cite{sun2015raptor} and how these can be circumvented~\cite{starov2016measuring},~\cite{nithyanand2016holding}~\cite{sun2017counter}. There are three key differences between the ToR relay network and the Bitcoin network that changes the spectrum of possible attacks and countermeasures. First, in order to protect the Bitcoin system we need to keep the network connected as opposed to preserving the privacy of every single connection for ToR. As such, we can use redundancy to protect Bitcoin clients, by connecting them to  multiple \system relays such that there is no AS that can effectively hijack all connections.
Second, counter-measures against routing attacks on ToR are limited to avoiding routes that might be affected by BGP hijacks, while we structure \system to avoid the chance of an attacker to be able to divert it in the first place. This is possible because Bitcoin clients have no preference with respect to who to connect to as they can get the same information from almost any peer. Third, countermeasures against routing attacks on ToR do no deal with the case that the client itself is hijacked.

\myitem{Multicast protocols} Mbone~\cite{eriksson1994mbone}, was designed to multicast live videos and music streams in the Internet, where many routers do not support IP multicast. Using tunnelling,  Mbone traffic can stay under the radar of those routers. Despite its novelty and usefulness, this network does not take into consideration whether the used paths can be hijacked and does not deal with maliciously increased load. Finally, systems such Splitstream~\cite{castro2003splitstream} that aim to reduce the load per node, require a fixed set of participants and a certain structure among them which would limit the openness of our network (regular clients cannot easily come and go).

\section{Conclusion}
\label{sec:conclusion}

We presented \system, a relay network aimed at securing Bitcoin against routing
attacks. The key insight behind \system is to position the relay nodes in
secured locations, preventing AS-level attackers from diverting intra-relay
communications and reducing their ability to divert traffic destined to the
relay clients. To protect the nodes themselves, \system leverages a
hardware/software co-design (leveraging programmable data planes) to
perform most of the relay operations in hardware. We fully implemented \system
and demonstrated its effectiveness in protecting Bitcoin, with as little as 6
relay nodes.




\bibliographystyle{IEEEtranS}
\bibliography{IEEEabrv,paper}

\appendix

\section{Appendix}
\label{sec:appendix}

\subsection{Algorithms}
\label{app:algos}
Below we include the pseudocode of the two main algorithms described in the paper in Section~\ref{subsec:algo1} and Section~\ref{subsec:algo2} respectively.

\begin{algorithm}
		\begin{algorithmic}[1]	

		\Function{LocateRelays}{$C, C\_scens, N, k$}\newline
		\Comment{$C\_scens$ scenarios that each relay protects against}\newline
		\Comment{$N$ number of relays to deploy}\newline
		\Comment{$k$ desired connectivity}

		\State $R$ $\gets$ $\{\}$ 
		\Comment{Relays to be deployed}
		\State $R\_scens$ $\gets$ $\{\}$ 
		\Comment{scenarios R protects against}
		
		\While{R.length < N}
			\State $Cs$ $\gets$ $\{c:~c\in C \setminus R\ s.t. G[R\cup c] \text{is k-connected}\}$
			\State best\_r $\gets$ 
			$FindNext$ ($Cs,$$C\_scens$,$R\_scens$)
			\State $R.add(best\_r)$
		\EndWhile
		\State \textbf{return} \text{R}
		\EndFunction	
		\Function{FindNext}{$Cs,C\_scens, R\_scens$}

	\State best\_r $\gets$ None
	\State best\_scens $\gets$ $\{\}$
	\State best\_effect $\gets$ MAX
	\For {r in Cs}
		\State tmp\_scens $\gets$ R\_scens $\cup$ C\_scens[r]
		\If {R\_scens.effect < best\_effect}
			\State  best\_scens $\gets$ tmp\_scens
			\State  best\_effect $\gets$ R\_scens.effect
			\State  best\_r $\gets$ r	
		\EndIf 
	\EndFor

	\State \textbf{return} $best\_r$
	\EndFunction
\end{algorithmic}	
	\caption{Algorithm to find the best set of relays to connect. }
	\label{algo:algo1}
\end{algorithm}

\begin{algorithm}
	\begin{algorithmic}[1]
	\Function{MorePreferred}{$pathA, pathB$}
	\State typeSeqA $\gets$ path\_type(pathA)
	\State typeSeqB $\gets$ path\_type(pathB)

	\While {pathA \& pathB \& hopA.pick==hopB.pick}
		\State hopA $\gets$ pathA.pop()
		\State hopB $\gets$ pathB.pop()
		\State typeA $\gets$ typeSeqB.pop()
		\State typeB $\gets$ typeSeqB.pop()
	\EndWhile
		
			\If {$typeA \neq typeB$}
				\Switch{$(typeA, typeB)$}
					\Case{(customer, peer)}
					\State \textbf{return} \text{0}
					\EndCase
					\Case{(customer, provider)}
					\State \textbf{return} \text{0}
					\EndCase
					\Case{(peer, provider)}
					\State \textbf{return} \text{0}
					\EndCase
					\Case{(peer, customer)}
					\State \textbf{return} \text{1}
					\EndCase
					\Case{(provider, customer)}
					\State \textbf{return} \text{0}
					\EndCase
					\Case{(provider, peer)}
					\State \textbf{return} \text{1}
					\EndCase
				\EndSwitch
		\Else 
			\If {$len(pathA) = len(pathB)$}
				\State \textbf{return} \text{1}
				\Comment{In case of a tie, we prefer path B.}
			\Else
				\State \textbf{return} \text{argmin(len(pathA), len(pathB))}
			\EndIf
		\EndIf 
	\EndFunction

	\end{algorithmic}
	\caption{Compare two paths based on preference.}
	\label{algo:algo2}
\end{algorithm}

\subsection{Results with ties against the attacker}
\label{app:graphs}
Although \system significantly improves the security of Bitcoin against routing attacks the exact performance depends on the on the tie-breaking decisions, namely which path is chosen in cases that the competing routes are equivalent economically and length-wise. In Section~\ref{sec:node_eval}, we assumed that the tie always breaks for the attacker. In the following, we include the same result only now assuming that the tie-breaking favors the legitimate destination.

\begin{figure}[h]
 \centering
 \includegraphics[width=.9\columnwidth]{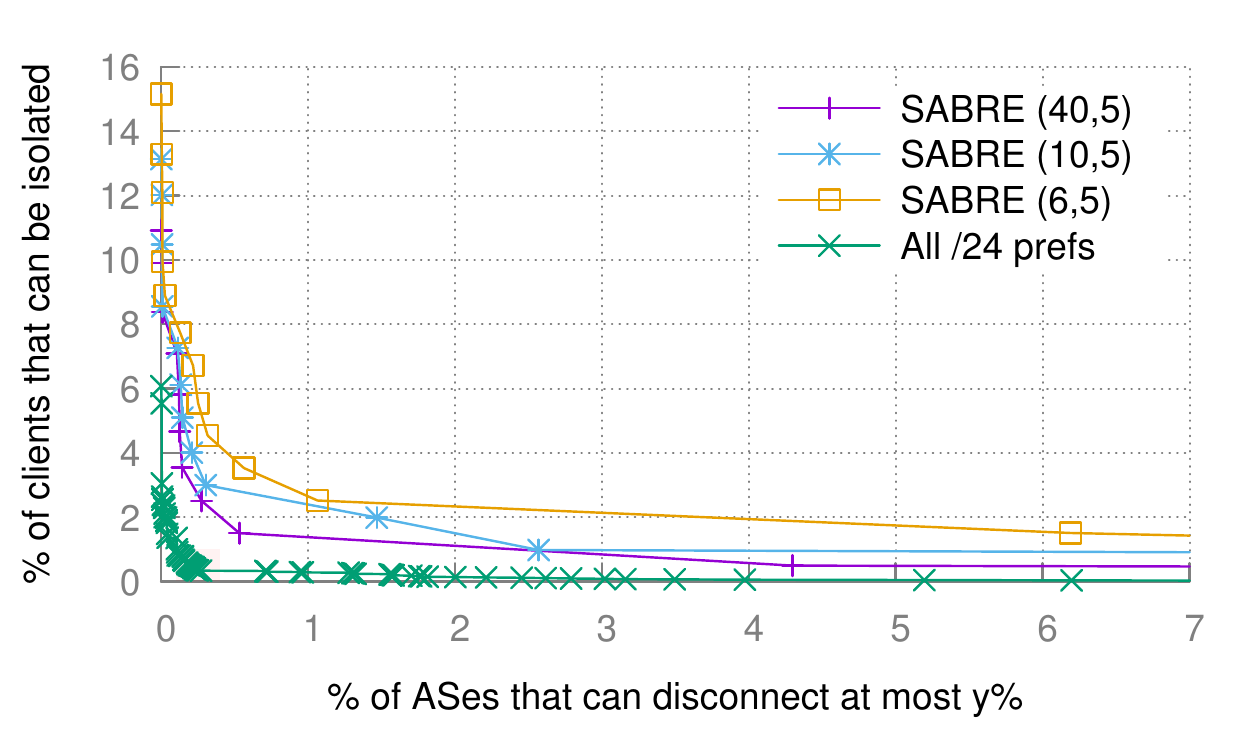}
 \caption{When tie breaks in favor of the legitimate destinations: a \system of only 6 relays that are fully connected prevents all attackers from isolating more than 16\%.}
 \label{fig:app1}
\end{figure}

\begin{figure}[h]
 \centering
 \includegraphics[width=.9\columnwidth]{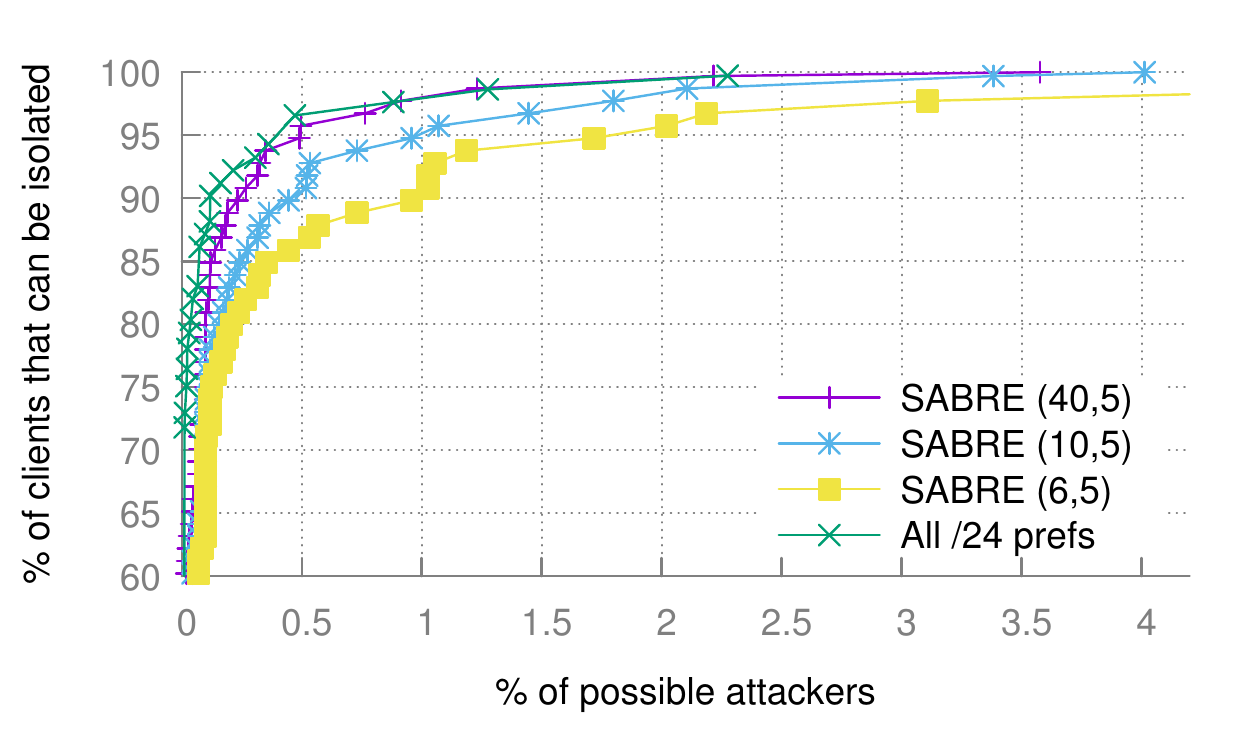}
 \caption{When tie breaks in favor of the legitimate destinations: a \system of 10 5-connected relays protects 95\% of the clients from 99.5\% of the AS level adversaries.}
 \label{fig:app2}
\end{figure}

\begin{figure}[h]
 \centering
 \includegraphics[width=.9\columnwidth]{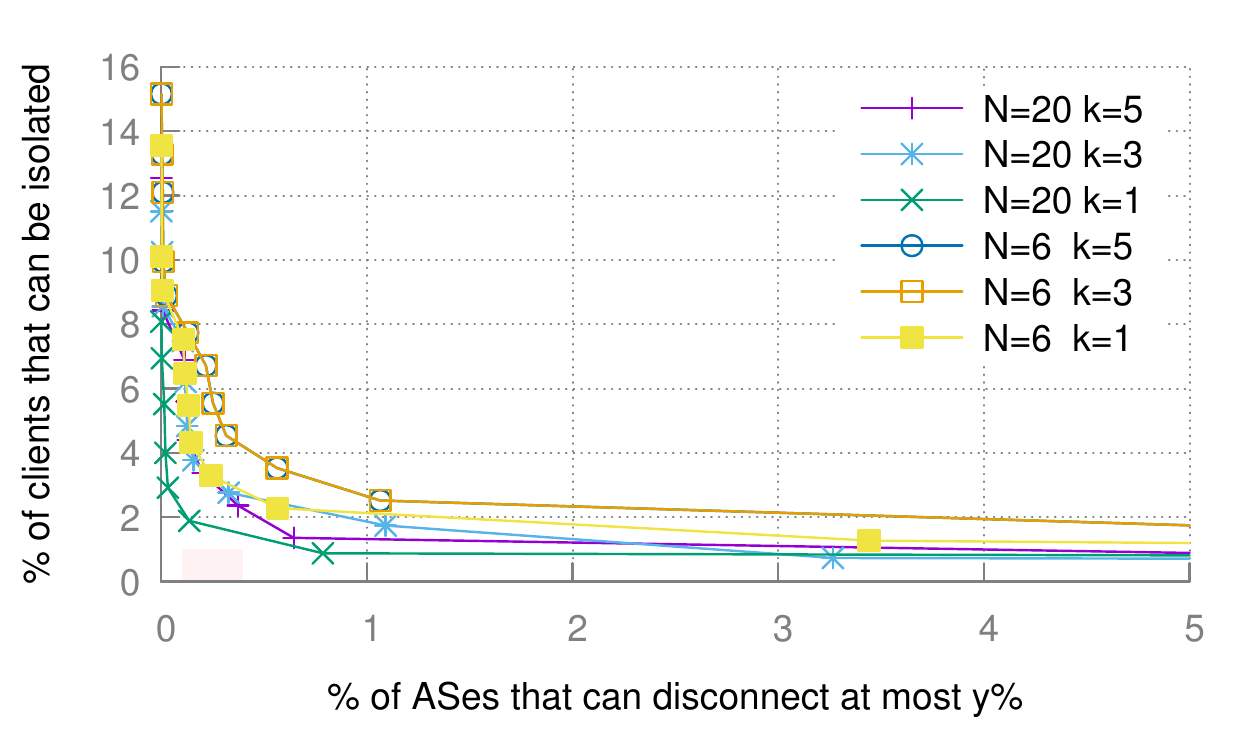}
 \caption{When tie breaks in favor of the legitimate destinations: the largest possible partition by any attacker is 14\% for a \system of 6 relays that is 5-connected.}
 \label{fig:app4}
\end{figure}

\begin{figure}[h]
 \centering
 \includegraphics[width=.9\columnwidth]{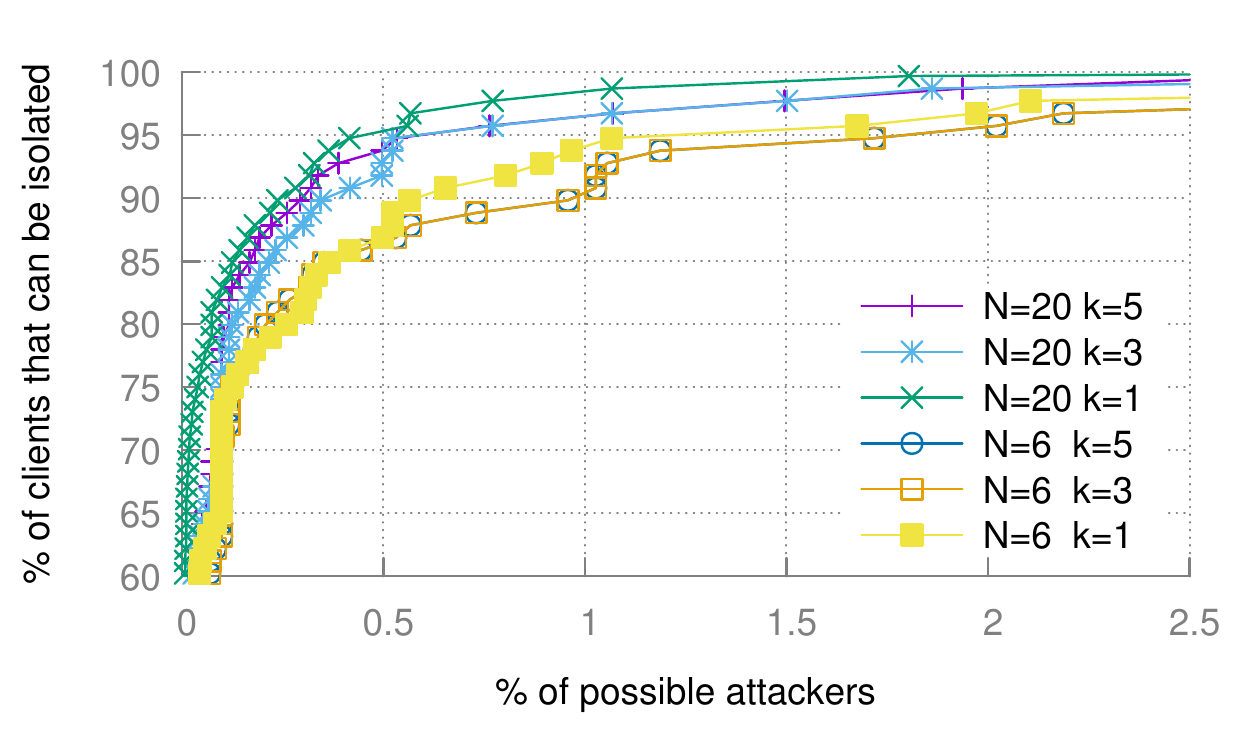}
 \caption{When tie breaks in favor of the legitimate destinations: a \system of 20  relays that are 1-connected can secure 100\% of the clients against more than 98\% attackers.}
 \label{fig:app3}
\end{figure}

\end{document}